\documentclass[aps,twocolumn,showpacs
,floatfix]{revtex4}
\usepackage{natbib}
\usepackage{amssymb,amsbsy,amsmath,amsfonts}
\usepackage{graphicx}

\newcommand{\SU}{{\rm SU}}
\newcommand{\U}{{\rm U}}

\newcommand{\comments}[1]{}

\begin{document}

\title{$\SU(6)\supset \SU(3) \otimes \SU(2)$ and $\SU(8)\supset \SU(4) \otimes
  \SU(2)$ Clebsch-Gordan coefficients}

\author{C. Garcia-Recio}
%\email{g_recio@ugr.es}
\author{L.L. Salcedo}
%\email{salcedo@ugr.es}

\affiliation{ Departamento de F{\'\i}sica At\'omica, Molecular y
Nuclear, Universidad de Granada, E-18071 Granada, Spain }

%\date{\today} 

\begin{abstract}
Tables of scalar factors are presented for $\mathbf{63}\otimes\mathbf{63}$ and
$\mathbf{120}\otimes\mathbf{63}$ in $\SU(8)\supset\SU(4)\otimes\SU(2)$, and
for $\mathbf{35}\otimes\mathbf{35}$ and $\mathbf{56}\otimes\mathbf{35}$ in
$\SU(6)\supset\SU(3)\otimes\SU(2)$. Related tables for $\SU(4)\supset
\SU(3)\otimes \U(1)$ and $\SU(3)\supset \SU(2) \otimes \U(1)$ are also
provided so that the Clebsch-Gordan coefficients can be completely
reconstructed. These are suitable to study meson-meson and baryon-meson within
a spin-flavor symmetric scheme.
\end{abstract}



\maketitle

\tableofcontents

\section{Introduction}

After the original SU(3) flavor symmetry, SU(6) spin-flavor symmetry was soon
introduced to give rise to the quark model
\cite{Gursey:1992dc,Pais:1964,Sakita:1964qq}, reaching considerable
phenomenological success. (See reviews in
\cite{Hey:1982aj,Capstick:2000qj,Fujiwara:2006yh}.)  It was shown early that
chiral symmetry, a powerful tool to extract information from QCD
\cite{Weinberg:1966kf,Tomozawa:1966jm,Gasser:1984gg}, and spin-flavor symmetry
were not in conflict \cite{Caldi:1975tx}. Recently, spin-flavor has reappeared
as a natural classification symmetry of baryons within the large $N_c$
approach to QCD \cite{Manohar:1998xv}, see
e.g. \cite{Goity:2002pu,Cherman:2009fh}. On the other hand, in the framework
of unitarized chiral models \cite{Truong:1988zp,Dobado:1989qm,
  Kaiser:1995eg,Oller:1997ng}, spin-flavor symmetry has found an important
role not only for three flavors but also for four
\cite{GarciaRecio:2005hy,GarciaRecio:2006wb,Toki:2007ab,GarciaRecio:2008dp,Gamermann:2010zz,GarciaRecio:2010vt,GarciaRecio:2010ki}. The
reason is that spin-flavor symmetry, in its chiral version, can naturally
accommodate the chiral and heavy quark symmetries of QCD. Although the
underlying $\SU(4)$ flavor symmetry is severely broken in the kinematics
through the quark masses, based on experience on three flavors, it is expected
that the breaking should be mild at the level of interaction amplitudes.

In view of this renewed interest in spin-flavor symmetry, we have undertaken a
study of the reduction of the spin-flavor group $\SU(8)$ under
$\SU(4)\otimes\SU(2)$, as required in the description of meson-meson and
baryon-meson interactions. Concretely we compute the scalar factors (also
called singlet factors) of the reduction $\SU(8)\supset\SU(4)\otimes\SU(2)$
for the products $\mathbf{63}\otimes\mathbf{63}$ (meson-meson) and
$\mathbf{120}\otimes\mathbf{63}$ (baryon-meson). These have not been yet
computed in the literature. It is true that, following the observation in
\cite{Chen:1981,Somers:1983en}, all scalar factors and Clebsch-Gordan
coefficients of special unitary groups follow from those of the symmetric
group, $S_n$, however, the irreducible representations of $\SU(8)$ considered
here would require large values of $n$ for which no results are readily
available.

The prescription of \cite{deSwart:1963gc} is widely used to define standard
bases of irreducible representations of $\SU(3)$
\cite{Amsler:2008zzb,Kaeding:1995vq}, however, for four or more flavors the
natural prescription is that of \cite{Baird:1964zm} (see,
e.g. \cite{Rabl:1975zy}). Previous calculations of scalar factors of
$\SU(6)\supset\SU(3)\otimes\SU(2)$ have been done using \cite{deSwart:1963gc},
for instance \cite{Cook65} (see also,
\cite{Carter:1965,Machacek:1975vc,Strottman:1979et,So:1979nw}), so we have
also computed the corresponding $\SU(6)\supset\SU(3)\otimes\SU(2)$ scalar
factors for $\mathbf{35}\otimes\mathbf{35}$ and
$\mathbf{56}\otimes\mathbf{35}$ with the prescription \cite{Baird:1964zm}.  We
reproduce the results in \cite{Cook65}, up to signs due to the different
choice of standard bases, and also differ in the breaking of degeneracies.

We present also the related $\SU(4)\supset\SU(3)\otimes\U(1)$ and
$\SU(3)\supset\SU(2)\otimes\U(1)$ scalar factors, always within the
prescription \cite{Baird:1964zm}. We check the results in
\cite{Rabl:1975zy}\footnote{We only differ in the value of $\xi$ (see
  Eq.~(\ref{eq:18}) below) for the two states $\mathbf{4^*}$ in
  $\mathbf{20}\otimes\mathbf{15}$ for $\SU(4)$.} and extend them, since we
need the product $\mathbf{20^\prime}\otimes\mathbf{15}$ in $\SU(4)$ which was
not computed there, along with the corresponding new
$\SU(3)\supset\SU(2)\otimes\U(1)$ scalar factors required.

In section \ref{sec:2} notation is introduced, in section \ref{sec:3} we
explain how to use the tables and in section \ref{sec:4} the method of
construction of the tables and the way the phases have been fixed is
described. The scalar factors are displayed in the appendices (see table of
contents).

\section{Representation reductions and Clebsch-Gordan series}
\label{sec:2}

For the groups $\SU(8)$ and $\SU(6)$ we consider the following reductions
\begin{eqnarray}
&&
\SU_{\rm sf}(8) \supset \SU_{\rm f}(4) \otimes \SU_{\rm J}(2)
,
\\ && 
\SU_{\rm sf}(6) \supset \SU_{\rm f}(3) \otimes \SU_{\rm J}(2)
,
\end{eqnarray}
with
\begin{eqnarray}
&&
\SU_{\rm f}(4) \supset \SU_{\rm f}(3) \otimes \U_{\rm C}(1)
,
\label{eq:2}
\\ && 
\SU_{\rm f}(3) \supset \SU_{\rm I}(2) \otimes \U_{\rm Y}(1)
\label{eq:3}
,\\ && 
\SU_{\rm I}(2) \supset \U_{\rm I_z}(1)
\label{eq:4}
,\\ && 
\SU_{\rm J}(2) \supset \U_{\rm J_z}(1)
\label{eq:5}
.
\end{eqnarray}
The labels sf, f, $J$, $C$, $I$, $Y$, $I_z$ and $J_z$ refer to spin-flavor,
flavor, spin, charm, isospin, hypercharge, third component of isospin and
third component of spin, respectively. In what follows we shall often drop
these labels in the groups.

To be concrete, we often refer to $\SU(8)$ although the following remarks
apply to the other groups as well with obvious modifications. Let the vector
space ${\mathcal H}_R$ carry an irreducible representation (irrep) $R$ of
$\SU(8)$, e.g. $\mathbf{63}$. This space admits an orthonormal basis,
$|R;\mu,\gamma,\zeta\rangle$, adapted to the reduction chains above. The label
$\mu$ denotes the $\SU(4)\otimes\SU(2)$ irrep, e.g. $\mathbf{15_1}$, where
$\mathbf{15}$ refers to $\SU(4)$ and the subindex $\mathbf{1}$ refers to
$2J+1$, with $J=0$.  $\gamma$ is the degeneracy label needed to distinguish
the various irreps $\mu$ appearing in the reduction of $R$ under 
$\SU(4) \otimes \SU(2)$. Finally, $\zeta$ denotes the remaining
quantum numbers $(\nu,I,J,I_z,J_z,Y,C)$, where $\nu$ denotes the $\SU(3)$
irrep, e.g., $\mathbf{8}$. The range of values of the label $\gamma$ depends
on $R$ and $\mu$. That of $\zeta$ depends on $\mu$.

Having two $\SU(8)$ irreducible spaces, ${\mathcal H}_{R_1}$ and ${\mathcal
  H}_{R_2}$, their tensor product can be decomposed under $\SU(8)$ as
\begin{equation}
{\mathcal H}_{R_1}\otimes {\mathcal H}_{R_2}
= \bigoplus_{R,\sigma} {\mathcal H}_{R,\sigma}
.
\end{equation}

The label $\sigma$ denotes the degeneracy label of the irrep $R$ in the
Clebsch-Gordan (CG) series of $R_1\otimes R_2$ and its range of values depends
on $R_1$, $R_2$ and $R$. The CG coefficients then relate the ``uncoupled'' and
``SU(8)-coupled'' bases as:
\begin{widetext}
\begin{eqnarray}
|R_1,R_2;R,\sigma;\mu,\gamma,\zeta\rangle
&=&
\sum_{\mu_1,\gamma_1,\zeta_1,\mu_2,\gamma_2,\zeta_2}
\left(
\begin{tabular}{ccc}
$R_1$ & $R_2$ & $R \sigma$ \\
$\mu_1\gamma_1\zeta_1$ & $\mu_2\gamma_2\zeta_2$ & $\mu\gamma\zeta$ \\
\end{tabular}
\right)
|R_1;\mu_1,\gamma_1,\zeta_1;R_2;\mu_2,\gamma_2,\zeta_2\rangle
,
\label{eq:7}
\\
|R_1;\mu_1,\gamma_1,\zeta_1;R_2;\mu_2,\gamma_2,\zeta_2\rangle
&=&
\sum_{R,\sigma,\mu\gamma,\zeta}
\left(
\begin{tabular}{ccc}
$R_1$ & $R_2$ & $R\sigma$ \\
$\mu_1\gamma_1\zeta_1$ & $\mu_2\gamma_2\zeta_2$ & $\mu\gamma\zeta$ \\
\end{tabular}
\right)
|R_1,R_2;R,\sigma;\mu,\gamma,\zeta\rangle
.
\end{eqnarray}
\end{widetext}
The symbols between large parenthesis denote the CG coefficients.

The CG coefficients pick up a plus or minus sign, $\xi$, under exchange of the
states 1 and 2,
\begin{equation}
\left(
\begin{tabular}{ccc}
$R_2$ & $R_1$ & $R \sigma$ \\
$\mu_2\gamma_2\zeta_2$ & $\mu_1\gamma_1\zeta_1$ & $\mu\gamma\zeta$ \\
\end{tabular}
\right)
=
\xi
\left(
\begin{tabular}{ccc}
$R_1$ & $R_2$ & $R \sigma$ \\
$\mu_1\gamma_1\zeta_1$ & $\mu_2\gamma_2\zeta_2$ & $\mu\gamma\zeta$ \\
\end{tabular}
\right)
.
\label{eq:18}
\end{equation}
The phase $\xi$ depends only on $R_1$, $R_2$, $R$, $\sigma$, $\mu$ and
$\gamma$.

On the other hand, the uncoupled basis can be coupled under
$\SU(4)\otimes\SU(2)$ 

\begin{widetext}
\begin{eqnarray}
|R_1,\mu_1,\gamma_1;R_2,\mu_2,\gamma_2;\mu,\gamma^\prime,\zeta\rangle
&=&
\sum_{\zeta_1,\zeta_2}
\left(
\begin{tabular}{ccc}
$\mu_1$ & $\mu_2$ & $\mu\gamma^\prime$ \\
$\zeta_1$ & $\zeta_2$ & $\zeta$ \\
\end{tabular}
\right)
|R_1;\mu_1,\gamma_1,\zeta_1;R_2;\mu_2,\gamma_2,\zeta_2\rangle
,
\label{eq:9}
\\
|R_1;\mu_1,\gamma_1,\zeta_1;R_2;\mu_2,\gamma_2,\zeta_2\rangle
&=&
\sum_{\mu,\gamma^\prime,\zeta}
\left(
\begin{tabular}{ccc}
$\mu_1$ & $\mu_2$ & $\mu\gamma^\prime$ \\
$\zeta_1$ & $\zeta_2$ & $\zeta$ \\
\end{tabular}
\right)
|R_1,\mu_1,\gamma_1;R_2,\mu_2,\gamma_2;\mu,\gamma^\prime,\zeta\rangle
,
\label{eq:10}
\end{eqnarray}
\end{widetext}
where $\gamma^\prime$ is the degeneracy label of the irrep $\mu$ in the CG
series of $\mu_1\otimes\mu_2$. Hence its range of values depends on $\mu_1$,
$\mu_2$ and $\mu$, but not on $\gamma_1$ and $\gamma_2$. The CG coefficients
in Eqs.~(\ref{eq:9},\ref{eq:10}) are the product of the CG coefficients of the
reductions chains $\SU(4)\supset\SU(3)\otimes \U(1)$ and $\SU(2)\supset
\U(1)$.

The relation between the $\SU(8)$-coupled and $\SU(4)\otimes\SU(2)$-coupled
basis provides the $\SU(8)\supset\SU(4)\otimes\SU(2)$ scalar factors (SF):
\begin{widetext}
\begin{eqnarray}
|R_1,R_2;R,\sigma;\mu,\gamma,\zeta\rangle
&=&
\sum_{\mu_1,\gamma_1,\mu_2,\gamma_2,\gamma^\prime}
\left(
\begin{tabular}{cc|c}
$R_1$ & $R_2$ &  $R\sigma$ \\
$\mu_1\gamma_1$ & $\mu_2\gamma_2$ & $\mu\gamma\gamma^\prime$ \\
\end{tabular}
\right)
|R_1,\mu_1,\gamma_1;R_2,\mu_2,\gamma_2;\mu,\gamma^\prime,\zeta\rangle
,
\label{eq:11}
\\
|R_1,\mu_1,\gamma_1;R_2,\mu_2,\gamma_2;\mu,\gamma^\prime,\zeta\rangle
&=&
\sum_{R,\sigma,\gamma}
\left(
\begin{tabular}{cc|c}
$R_1$ & $R_2$ & $R\sigma$ \\
$\mu_1\gamma_1$ & $\mu_2\gamma_2$ & $\mu\gamma\gamma^\prime$ \\
\end{tabular}
\right)
|R_1,R_2;R,\sigma;\mu,\gamma,\zeta\rangle
.
\label{eq:12}
\end{eqnarray}
\end{widetext}
The symbols between large parenthesis and vertical bar denote the SF.

The degeneracy labels are redundant when there is no degeneracy.  Following
the standard practice, throughout this work, in those cases the label is
omitted.  This refers either to degeneracy in the irrep reduction ($\gamma$)
or in the CG series ($\sigma$, $\gamma^\prime$). As degeneracy labels we use
$s$, $a$ and $b$ (or $\mathbf{s}$, $\mathbf{a}$, $\mathbf{b}$), in this
order. Three is the highest degeneracy encountered.  In the general case, the
labels $s$ and $a$ do not imply symmetry of the representation.  When there is
symmetry (e.g. in $\mathbf{63}\otimes\mathbf{63}$) we take the symmetric
representation to be the first one ($s$) and the antisymmetric representation
to be second one ($a$). In those cases the phase $\xi=\pm 1$ for $s$ and $a$
respectively.

There are completely analogous relations for CG and SF of $\SU(6)\supset
\SU(3)\otimes\SU(2)$, $\SU(4)\supset \SU(3)\otimes\U(1)$, and $\SU(3)\supset
\SU(2)\otimes\U(1)$, and similar comments apply.  Because the reduction chains
in Eqs.~(\ref{eq:2}-\ref{eq:5}) are canonical no degeneracy label $\gamma$ is
needed for $\SU(4)$ and $\SU(3)$.

\section{Explanation of the tables}
\label{sec:3}

For $\SU(8)$ we consider the following CG series
\begin{eqnarray}
\mathbf{63}\otimes\mathbf{63} &=&
\mathbf{1}\oplus\mathbf{63_s}\oplus\mathbf{720}\oplus\mathbf{1232}\oplus\mathbf{63_a}\oplus\mathbf{945}\oplus\mathbf{945^*},
\nonumber \\
\mathbf{120}\otimes\mathbf{63} &=&
\mathbf{120}\oplus\mathbf{168}\oplus\mathbf{2520}\oplus\mathbf{4752},
\end{eqnarray}
and for $\SU(6)$ 
\begin{eqnarray}
\mathbf{35}\otimes\mathbf{35} &=&
\mathbf{1}\oplus\mathbf{35_s}\oplus\mathbf{189}\oplus\mathbf{405}\oplus\mathbf{35_a}\oplus\mathbf{280}\oplus\mathbf{280^*},
\nonumber \\
\mathbf{56}\otimes\mathbf{35} &=&
\mathbf{56}\oplus\mathbf{70}\oplus\mathbf{700}\oplus\mathbf{1134}.
\end{eqnarray}
The irreps are labeled by their dimension. The labels used and the
corresponding Young tableaux are listed in table \ref{tab:1}.

\begin{table}
\begin{center}
\caption{List of Young tableaux and irrep labels. }
\label{tab:1}
\begin{tabular}{lll}
Group & irrep label & Young tableau  \\
\hline
\hline
SU(8) 
  & $\mathbf{1}$ & $[~] $\\
  & $\mathbf{8}$ & $[1] $\\
  & $\mathbf{8^*}$ & $[1^7] $\\
  & $\mathbf{63}$ & $[2,1^6] $\\
  & $\mathbf{120}$ & $[3] $\\
  & $\mathbf{168}$ & $[2,1] $\\
  & $\mathbf{720}$ & $[2^2,1^4] $\\
  & $\mathbf{945}$ & $[3,1^5] $\\
  & $\mathbf{945^*}$ & $[3^2,2^5] $\\
  & $\mathbf{1232}$ & $[4,2^6] $\\
  & $\mathbf{2520}$ & $[5,1^6] $\\
  & $\mathbf{4752}$ & $[4,2,1^5] $\\
\hline
SU(6) 
  %& $\mathbf{1}$ & $[~] $\\
  & $\mathbf{6}$ & $[1] $\\
  & $\mathbf{6^*}$ & $[1^5] $\\
  & $\mathbf{35}$ & $[2,1^4] $\\
  & $\mathbf{56}$ & $[3] $\\
  & $\mathbf{70}$ & $[2,1] $\\
  & $\mathbf{189}$ & $[2^2,1^2] $\\
  & $\mathbf{280}$ & $[3,1^3] $\\
  & $\mathbf{280^*}$ & $[3^2,2^3] $\\
  & $\mathbf{405}$ & $[4,2^4] $\\
  & $\mathbf{700}$ & $[5,1^4] $\\
  & $\mathbf{1134}$ & $[4,2,1^3] $\\
\hline
SU(4) %& $\mathbf{1}$ & $[~] $\\
  & $\mathbf{4}$ & $[1] $\\
  & $\mathbf{4^*}$ & $[1^3] $\\
  & $\mathbf{15}$ & $[2,1^2] $\\
  & $\mathbf{20}$ & $[2,1] $\\
  & $\mathbf{20^\prime}$ & $[3] $\\
  & $\mathbf{20^{\prime\prime}}$ & $[2^2] $\\
  & $\mathbf{36^*}$ & $[3,2^2] $\\
  & $\mathbf{45}$ & $[3,1] $\\
  & $\mathbf{45^*}$ & $[3^2,2] $\\
  & $\mathbf{60^*}$ & $[3^2,1] $\\
  & $\mathbf{84}$ & $[4,2^2] $\\
  & $\mathbf{120}$ & $[5,1^2] $\\
  & $\mathbf{140}$ & $[4,2,1] $\\
\hline
SU(3) %& $\mathbf{1}$ & $[~] $\\
  & $\mathbf{3}$ & $[1] $\\
  & $\mathbf{3^*}$ & $[1^2] $\\
  & $\mathbf{6}$ & $[2] $\\
  & $\mathbf{6^*}$ & $[2^2] $\\
  & $\mathbf{8}$ & $[2,1] $\\
  & $\mathbf{10}$ & $[3] $\\
  & $\mathbf{10^*}$ & $[3^2] $\\
  & $\mathbf{15}$ & $[3,1] $\\
  & $\mathbf{15^*}$ & $[3,2] $\\
  & $\mathbf{15^\prime}$ & $[4] $\\
  & $\mathbf{24^*}$ & $[4,1] $\\
  & $\mathbf{27}$ & $[4,2] $\\
  & $\mathbf{35}$ & $[5,1] $\\
\hline
\end{tabular}
\end{center}
\end{table}

The tables of SF are presented in the appendices.  The related SU(4) and SU(3)
reductions are also included. The reduction of the irreps in the CG series is
obvious from the SF tables.  Nevertheless, for convenience we have collected
in table \ref{tab:2} these reductions for $\SU(8)\supset\SU(4)\otimes\SU(2)$
and for $\SU(6)\supset\SU(3)\otimes\SU(2)$.

\begin{table}
\begin{center}
\caption{Reduction of SU(8) and SU(6) irreps. }
\label{tab:2}
\begin{tabular}{ll}
\hline
SU(8) & SU(4) $\otimes$ SU(2)  \\
\hline
$\mathbf{63}$ & 
  $ \mathbf{1_3} $ 
  $ \mathbf{15_1} $ 
  $ \mathbf{15_3} $ 
\\
$\mathbf{120}$ & 
  $ \mathbf{20_2} $ 
  $ \mathbf{20^\prime_4} $ 
\\
$\mathbf{168}$ & 
  $ \mathbf{4^*_2} $ 
  $ \mathbf{20_2} $ 
  $ \mathbf{20_4} $ 
  $ \mathbf{20^\prime_2} $ 
\\
$\mathbf{720}$ & 
  $ \mathbf{1_1} $ 
  $ \mathbf{1_5} $ 
  $ \mathbf{15_1} $ 
  $ \mathbf{15_{s,3}} $ 
  $ \mathbf{15_{a,3}} $ 
  $ \mathbf{15_5} $ 
  $ \mathbf{20^{\prime\prime}_1} $ 
  $ \mathbf{20^{\prime\prime}_3} $ 
  $ \mathbf{20^{\prime\prime}_5} $ 
  $ \mathbf{45_3} $ 
  $ \mathbf{45^*_3} $ 
  $ \mathbf{84_1} $ 
\\
$\mathbf{945}$ & 
  $ \mathbf{1_3} $ 
  $ \mathbf{15_1} $ 
  $ \mathbf{15_{s,3}} $ 
  $ \mathbf{15_{a,3}} $ 
  $ \mathbf{15_5} $ 
  $ \mathbf{20^{\prime\prime}_3} $ 
  $ \mathbf{45_1} $ 
  $ \mathbf{45_3} $ 
  $ \mathbf{45_5} $ 
  $ \mathbf{45^*_1} $ 
  $ \mathbf{84_3} $ 
\\
$\mathbf{945^*}$ & 
  $ \mathbf{1_3} $ 
  $ \mathbf{15_1} $ 
  $ \mathbf{15_{s,3}} $ 
  $ \mathbf{15_{a,3}} $ 
  $ \mathbf{15_5} $ 
  $ \mathbf{20^{\prime\prime}_3} $ 
  $ \mathbf{45_1} $ 
  $ \mathbf{45^*_1} $ 
  $ \mathbf{45^*_3} $ 
  $ \mathbf{45^*_5} $ 
  $ \mathbf{84_3} $ 
\\
$\mathbf{1232}$ & 
  $ \mathbf{1_1} $ 
  $ \mathbf{1_5} $ 
  $ \mathbf{15_1} $ 
  $ \mathbf{15_{s,3}} $ 
  $ \mathbf{15_{a,3}} $ 
  $ \mathbf{15_5} $ 
  $ \mathbf{20^{\prime\prime}_1} $ 
  $ \mathbf{45_3} $ 
  $ \mathbf{45^*_3} $ 
  $ \mathbf{84_1} $ 
  $ \mathbf{84_3} $ 
  $ \mathbf{84_5} $ 
\\
$\mathbf{2520}$ & 
  $ \mathbf{20_2} $ 
  $ \mathbf{20_4} $ 
  $ \mathbf{20^\prime_2} $ 
  $ \mathbf{20^\prime_4} $ 
  $ \mathbf{20^\prime_6} $ 
  $ \mathbf{60^*_2} $ 
  $ \mathbf{120_4} $ 
  $ \mathbf{120_6} $ 
  $ \mathbf{140_2} $ 
  $ \mathbf{140_4} $ 
\\
$\mathbf{4752}$ & 
  $ \mathbf{4^*_2} $ 
  $ \mathbf{4^*_4} $ 
  $ \mathbf{20_{s,2}} $ 
  $ \mathbf{20_{a,2}} $ 
  $ \mathbf{20_{b,2}} $ 
  $ \mathbf{20_{s,4}} $ 
  $ \mathbf{20_{a,4}} $ 
  $ \mathbf{20_{b,4}} $ 
  $ \mathbf{20_6} $
\\ &  
  $ \mathbf{20^\prime_{s,2}} $ 
  $ \mathbf{20^\prime_{a,2}} $ 
  $ \mathbf{20^\prime_{s,4}} $ 
  $ \mathbf{20^\prime_{a,4}} $ 
  $ \mathbf{20^\prime_6} $ 
  $ \mathbf{36^*_{s,2}} $ 
  $ \mathbf{36^*_{a,2}} $ 
  $ \mathbf{36^*_4} $ 
  $ \mathbf{60^*_2} $ 
  $ \mathbf{60^*_4} $ 
\\ &  
  $ \mathbf{120_2} $ 
  $ \mathbf{120_4} $ 
  $ \mathbf{140_{s,2}} $ 
  $ \mathbf{140_{a,2}} $ 
  $ \mathbf{140_{s,4}} $ 
  $ \mathbf{140_{a,4}} $ 
  $ \mathbf{140_6} $ 
\\
\hline
\hline
SU(6) & SU(3) $\otimes$ SU(2)  \\
\hline
$\mathbf{35}$ & 
  $ \mathbf{1_3} $ 
  $ \mathbf{8_1} $ 
  $ \mathbf{8_3} $ 
\\
$\mathbf{56}$ & 
  $ \mathbf{8_2} $ 
  $ \mathbf{10_4} $ 
\\
$\mathbf{70}$ & 
  $ \mathbf{1_2} $ 
  $ \mathbf{8_2} $ 
  $ \mathbf{8_4} $ 
  $ \mathbf{10_2} $ 
\\
$\mathbf{189}$ & 
  $ \mathbf{1_1} $ 
  $ \mathbf{1_5} $ 
  $ \mathbf{8_1} $ 
  $ \mathbf{8_{s,3}} $ 
  $ \mathbf{8_{a,3}} $ 
  $ \mathbf{8_5} $ 
  $ \mathbf{10_3} $ 
  $ \mathbf{10^*_3} $ 
  $ \mathbf{27_1} $ 
\\
$\mathbf{280}$ & 
  $ \mathbf{1_3} $ 
  $ \mathbf{8_1} $ 
  $ \mathbf{8_{s,3}} $ 
  $ \mathbf{8_{a,3}} $ 
  $ \mathbf{8_5} $ 
  $ \mathbf{10_1} $ 
  $ \mathbf{10_3} $ 
  $ \mathbf{10_5} $ 
  $ \mathbf{10^*_1} $ 
  $ \mathbf{27_3} $ 
\\
$\mathbf{280^*}$ & 
  $ \mathbf{1_3} $ 
  $ \mathbf{8_1} $ 
  $ \mathbf{8_{s,3}} $ 
  $ \mathbf{8_{a,3}} $ 
  $ \mathbf{8_5} $ 
  $ \mathbf{10_1} $ 
  $ \mathbf{10^*_1} $ 
  $ \mathbf{10^*_3} $ 
  $ \mathbf{10^*_5} $ 
  $ \mathbf{27_3} $ 
\\
$\mathbf{405}$ & 
  $ \mathbf{1_1} $ 
  $ \mathbf{1_5} $ 
  $ \mathbf{8_1} $ 
  $ \mathbf{8_{s,3}} $ 
  $ \mathbf{8_{a,3}} $ 
  $ \mathbf{8_5} $ 
  $ \mathbf{10_3} $ 
  $ \mathbf{10^*_3} $ 
  $ \mathbf{27_1} $ 
  $ \mathbf{27_3} $ 
  $ \mathbf{27_5} $ 
\\
$\mathbf{700}$ & 
  $ \mathbf{8_2} $ 
  $ \mathbf{8_4} $ 
  $ \mathbf{10_2} $ 
  $ \mathbf{10_4} $ 
  $ \mathbf{10_6} $ 
  $ \mathbf{10^*_2} $ 
  $ \mathbf{27_2} $ 
  $ \mathbf{27_4} $ 
  $ \mathbf{35_4} $ 
  $ \mathbf{35_6} $ 
\\
$\mathbf{1134}$ & 
  $ \mathbf{1_2} $ 
  $ \mathbf{1_4} $ 
  $ \mathbf{8_{s,2}} $ 
  $ \mathbf{8_{a,2}} $ 
  $ \mathbf{8_{b,2}} $ 
  $ \mathbf{8_{s,4}} $ 
  $ \mathbf{8_{a,4}} $ 
  $ \mathbf{8_{b,4}} $ 
  $ \mathbf{8_6} $ 
\\ &  
  $ \mathbf{10_{s,2}} $ 
  $ \mathbf{10_{a,2}} $ 
  $ \mathbf{10_{s,4}} $ 
  $ \mathbf{10_{a,4}} $ 
  $ \mathbf{10_6} $ 
  $ \mathbf{10^*_2} $ 
  $ \mathbf{10^*_4} $ 
\\ &  
  $ \mathbf{27_{s,2}} $ 
  $ \mathbf{27_{a,2}} $ 
  $ \mathbf{27_{s,4}} $ 
  $ \mathbf{27_{a,4}} $ 
  $ \mathbf{27_6} $ 
  $ \mathbf{35_2} $ 
  $ \mathbf{35_4} $ 
\\
\hline
\end{tabular}
\end{center}
\end{table}

Once again, to be concrete, we discuss the SU(8) case. The tables are
expressed in the form of equations, in terms of state vectors as in
Eq.~(\ref{eq:11}), i.e. a l.h.s with the $\SU(8)$-coupled stated and a
r.h.s. with the $\SU(4)\otimes\SU(2)$-coupled state. From such equations the
SF can be read off. The plus or minus sign between parenthesis displayed at
the left of the l.h.s. of the equation is the phase $\xi$ defined in
Eq.~(\ref{eq:18}). It is put there for the sake of presentation only (it is
not meant to multiply the vector).

Regarding the notation, as compared to Eq.~(\ref{eq:11}), in the tables,
redundant labels have been omitted.  Specifically, in the l.h.s. we omit
$R_1$, $R_2$ since they are explicited in the heading of the corresponding
subsection, and $\zeta$ is also omitted. The degeneracy labels $\sigma$ and
$\gamma$, if required, take the form of a subindex. E.g.,
$|\mathbf{63_a};\mathbf{15_3}\rangle$ ~ ($\sigma=a$), or
$|\mathbf{1232};\mathbf{15_{s,3}}\rangle$ ~ ($\gamma=s$).

In the r.h.s., $\gamma_1$ and $\gamma_2$ are not needed for the $\SU(8)$
representations $R_1$ and $R_2$ considered, namely, $\mathbf{120}$ or
$\mathbf{63}$.  The labels $\mu$ and $\zeta$ are also omitted ($\mu$ is
explicited in the l.h.s.). The irreps $(R_1,\mu_1)$ and $(R_2,\mu_2)$ are
represented by symbols of the lowest lying particles with those quantum
numbers. In each case, {\em the particle with highest weight is used}. E.g.,
$(\mathbf{120},\mathbf{20_2})$ is labeled by $\Sigma$ and
$(\mathbf{63},\mathbf{15_3})$ is labeled by $\rho$. The list of particle
symbols is displayed in Table~\ref{tab:3} and there the rows are ordered so
that they run from highest (top) to lowest weight (bottom).

\begin{table}
\begin{center}
\caption{List of particle symbols. }
\label{tab:3}
\begin{tabular}{cccccrrrr}
label & SU(8) & SU(6) & SU(4) & SU(3) & $I$ & $Y$ & $C$ & $J$ \\
\hline
$\Delta$
& $\mathbf{120}$
& $\mathbf{56}$
& $\mathbf{20^\prime}$
& $\mathbf{10}$
&$\frac{3}{2}$
&$1$
&$0$
&$\frac{3}{2}$
\\
$\Sigma^*$
& $\mathbf{120}$
& $\mathbf{56}$
& $\mathbf{20^\prime}$
& $\mathbf{10}$
&$1$
&$0$
&$0$
&$\frac{3}{2}$
\\
$\Xi^*$
& $\mathbf{120}$
& $\mathbf{56}$
& $\mathbf{20^\prime}$
& $\mathbf{10}$
&$\frac{1}{2}$
&$-1$
&$0$
&$\frac{3}{2}$
\\
$\Omega$
& $\mathbf{120}$
& $\mathbf{56}$
& $\mathbf{20^\prime}$
& $\mathbf{10}$
&$0$
&$-2$
&$0$
&$\frac{3}{2}$
\\
$\Sigma_c^*$
& $\mathbf{120}$
& $\mathbf{ }$
& $\mathbf{20^\prime}$
& $\mathbf{6}$
&$1$
&$\frac{2}{3}$
&$1$
&$\frac{3}{2}$
\\
$\Xi_c^*$
& $\mathbf{120}$
& $\mathbf{ }$
& $\mathbf{20^\prime}$
& $\mathbf{6}$
&$\frac{1}{2}$
&$-\frac{1}{3}$
&$1$
&$\frac{3}{2}$
\\
$\Omega_c^*$
& $\mathbf{120}$
& $\mathbf{ }$
& $\mathbf{20^\prime}$
& $\mathbf{6}$
&$0$
&$-\frac{4}{3}$
&$1$
&$\frac{3}{2}$
\\
$\Xi_{cc}^*$
& $\mathbf{120}$
& $\mathbf{ }$
& $\mathbf{20^\prime}$
& $\mathbf{3}$
&$\frac{1}{2}$
&$\frac{1}{3}$
&$2$
&$\frac{3}{2}$
\\
$\Omega_{cc}^*$
& $\mathbf{120}$
& $\mathbf{ }$
& $\mathbf{20^\prime}$
& $\mathbf{3}$
&$0$
&$-\frac{2}{3}$
&$2$
&$\frac{3}{2}$
\\
$\Omega_{ccc}$
& $\mathbf{120}$
& $\mathbf{ }$
& $\mathbf{20^\prime}$
& $\mathbf{1}$
&$0$
&$0$
&$3$
&$\frac{3}{2}$
\\
$\Sigma$
& $\mathbf{120}$
& $\mathbf{56}$
& $\mathbf{20}$
& $\mathbf{8}$
&$1$
&$0$
&$0$
&$\frac{1}{2}$
\\
$N$
& $\mathbf{120}$
& $\mathbf{56}$
& $\mathbf{20}$
& $\mathbf{8}$
&$\frac{1}{2}$
&$1$
&$0$
&$\frac{1}{2}$
\\
$\Xi$
& $\mathbf{120}$
& $\mathbf{56}$
& $\mathbf{20}$
& $\mathbf{8}$
&$\frac{1}{2}$
&$-1$
&$0$
&$\frac{1}{2}$
\\
$\Lambda$
& $\mathbf{120}$
& $\mathbf{56}$
& $\mathbf{20}$
& $\mathbf{8}$
&$0$
&$0$
&$0$
&$\frac{1}{2}$
\\
$\Sigma_c$
& $\mathbf{120}$
& $\mathbf{ }$
& $\mathbf{20}$
& $\mathbf{6}$
&$1$
&$\frac{2}{3}$
&$1$
&$\frac{1}{2}$
\\
$\Xi_c^\prime$
& $\mathbf{120}$
& $\mathbf{ }$
& $\mathbf{20}$
& $\mathbf{6}$
&$\frac{1}{2}$
&$-\frac{1}{3}$
&$1$
&$\frac{1}{2}$
\\
$\Omega_c$
& $\mathbf{120}$
& $\mathbf{ }$
& $\mathbf{20}$
& $\mathbf{6}$
&$0$
&$-\frac{4}{3}$
&$1$
&$\frac{1}{2}$
\\
$\Xi_c$
& $\mathbf{120}$
& $\mathbf{ }$
& $\mathbf{20}$
& $\mathbf{3^*}$
&$\frac{1}{2}$
&$-\frac{1}{3}$
&$1$
&$\frac{1}{2}$
\\
$\Lambda_c$
& $\mathbf{120}$
& $\mathbf{ }$
& $\mathbf{20}$
& $\mathbf{3^*}$
&$0$
&$\frac{2}{3}$
&$1$
&$\frac{1}{2}$
\\
$\Xi_{cc}$
& $\mathbf{120}$
& $\mathbf{ }$
& $\mathbf{20}$
& $\mathbf{3}$
&$\frac{1}{2}$
&$\frac{1}{3}$
&$2$
&$\frac{1}{2}$
\\
$\Omega_{cc}$
& $\mathbf{120}$
& $\mathbf{ }$
& $\mathbf{20}$
& $\mathbf{3}$
&$0$
&$-\frac{2}{3}$
&$2$
&$\frac{1}{2}$
\\
$\rho$
& $\mathbf{63}$
& $\mathbf{35}$
& $\mathbf{15}$
& $\mathbf{8}$
&$1$
&$0$
&$0$
&$1$
\\
$K^*$
& $\mathbf{63}$
& $\mathbf{35}$
& $\mathbf{15}$
& $\mathbf{8}$
&$\frac{1}{2}$
&$1$
&$0$
&$1$
\\
${\bar K}^*$
& $\mathbf{63}$
& $\mathbf{35}$
& $\mathbf{15}$
& $\mathbf{8}$
&$\frac{1}{2}$
&$-1$
&$0$
&$1$
\\
$\omega_8$
& $\mathbf{63}$
& $\mathbf{35}$
& $\mathbf{15}$
& $\mathbf{8}$
&$0$
&$0$
&$0$
&$1$
\\
$D^*$
& $\mathbf{63}$
& $\mathbf{ }$
& $\mathbf{15}$
& $\mathbf{3^*}$
&$\frac{1}{2}$
&$-\frac{1}{3}$
&$1$
&$1$
\\
$D_s^*$
& $\mathbf{63}$
& $\mathbf{ }$
& $\mathbf{15}$
& $\mathbf{3^*}$
&$0$
&$\frac{2}{3}$
&$1$
&$1$
\\
${\bar D}^*$
& $\mathbf{63}$
& $\mathbf{ }$
& $\mathbf{15}$
& $\mathbf{3}$
&$\frac{1}{2}$
&$\frac{1}{3}$
&$-1$
&$1$
\\
${\bar D}_s^*$
& $\mathbf{63}$
& $\mathbf{ }$
& $\mathbf{15}$
& $\mathbf{3}$
&$0$
&$-\frac{2}{3}$
&$-1$
&$1$
\\
$\psi$
& $\mathbf{63}$
& $\mathbf{ }$
& $\mathbf{15}$
& $\mathbf{1}$
&$0$
&$0$
&$0$
&$1$
\\
$\omega_1$
& $\mathbf{63}$
& $\mathbf{35}$
& $\mathbf{1}$
& $\mathbf{1}$
&$0$
&$0$
&$0$
&$1$
\\
$\pi$
& $\mathbf{63}$
& $\mathbf{35}$
& $\mathbf{15}$
& $\mathbf{8}$
&$1$
&$0$
&$0$
&$0$
\\
$K$
& $\mathbf{63}$
& $\mathbf{35}$
& $\mathbf{15}$
& $\mathbf{8}$
&$\frac{1}{2}$
&$1$
&$0$
&$0$
\\
${\bar K}$
& $\mathbf{63}$
& $\mathbf{35}$
& $\mathbf{15}$
& $\mathbf{8}$
&$\frac{1}{2}$
&$-1$
&$0$
&$0$
\\
$\eta$
& $\mathbf{63}$
& $\mathbf{35}$
& $\mathbf{15}$
& $\mathbf{8}$
&$0$
&$0$
&$0$
&$0$
\\
$D$
& $\mathbf{63}$
& $\mathbf{ }$
& $\mathbf{15}$
& $\mathbf{3^*}$
&$\frac{1}{2}$
&$-\frac{1}{3}$
&$1$
&$0$
\\
$D_s$
& $\mathbf{63}$
& $\mathbf{ }$
& $\mathbf{15}$
& $\mathbf{3^*}$
&$0$
&$\frac{2}{3}$
&$1$
&$0$
\\
${\bar D}$
& $\mathbf{63}$
& $\mathbf{ }$
& $\mathbf{15}$
& $\mathbf{3}$
&$\frac{1}{2}$
&$\frac{1}{3}$
&$-1$
&$0$
\\
${\bar D}_s$
& $\mathbf{63}$
& $\mathbf{ }$
& $\mathbf{15}$
& $\mathbf{3}$
&$0$
&$-\frac{2}{3}$
&$-1$
&$0$
\\
$\eta_c$
& $\mathbf{63}$
& $\mathbf{ }$
& $\mathbf{15}$
& $\mathbf{1}$
&$0$
&$0$
&$0$
&$0$
\\
$\eta^\prime$
& $\mathbf{1}$
& $\mathbf{1}$
& $\mathbf{1}$
& $\mathbf{1}$
&$0$
&$0$
&$0$
&$0$
\\
\hline
\end{tabular}
\end{center}
\end{table}

Further, if the degeneracy label $\gamma^\prime$ is required, it is put as a
subindex. E.g. $|(\Sigma,\rho)_s\rangle$ for $\gamma^\prime=s$. As a final
notational convention, in the r.h.s, when $R_1=R_2$ we use a label $S$ or $A$
to indicate symmetrized or antisymmetrized states under exchange of particle
labels. E.g.,
\begin{eqnarray}
|(\rho,\pi)_a\rangle_S &=& 
\frac{1}{\sqrt{2}}|(\rho,\pi)_a\rangle +
\frac{1}{\sqrt{2}}|(\pi,\rho)_a\rangle,
\nonumber \\
|\rho,\omega_1\rangle_A &=& 
\frac{1}{\sqrt{2}}|\rho,\omega_1\rangle 
-\frac{1}{\sqrt{2}}|\omega_1,\rho\rangle
.
\end{eqnarray}

Hence, Eq.~(\ref{eq:11}) are written in the appendices as
\begin{equation}
(\xi)
~~
|R_\sigma;\mu_\gamma\rangle
=
\sum_{\mu_1,\mu_2,\gamma^\prime}
\left(
\begin{tabular}{cc|c}
$R_1$ & $R_2$ &  $R\sigma$ \\
$\mu_1$ & $\mu_2$ & $\mu\gamma\gamma^\prime$ \\
\end{tabular}
\right)
|(\mu_1,\mu_2)_{\gamma^\prime}\rangle
.
\end{equation}

The equations corresponding to Eq.~(\ref{eq:12}) are not given as they are
easily reconstructed from the equations provided (of the type Eq.~(\ref{eq:11})). E.g., from the tables for
$R_1\otimes R_2=\mathbf{63}\otimes\mathbf{63}$ with $\mu=\mathbf{15_5}$, one
finds, for $\mu_1=\mu_2=\mathbf{15_3}$,
\begin{equation}
|(\rho,\rho)_s\rangle
=
\sqrt{\frac{3}{4}}|\mathbf{720};\mathbf{15_5}\rangle
+
\sqrt{\frac{1}{4}}|\mathbf{1232};\mathbf{15_5}\rangle
.
\end{equation}
As explained, here $|(\rho,\rho)_s\rangle=
|\mathbf{15_3}\otimes\mathbf{15_3};\mathbf{15_{s,5}}\rangle$.

For $\SU(6)\supset\SU(3)\otimes\SU(2)$ everything is similar. The particle
symbols in the r.h.s. correspond now to multiplets $(R_1,\mu_1)$ and
$(R_2,\mu_2)$ of $(\SU(6),\SU(3)\otimes\SU(2))$, so for instance, $\Sigma$
stands now for $(\mathbf{56},\mathbf{8_2})$ and $\rho$ stands for
$(\mathbf{35},\mathbf{8_3})$.

For $\SU(4)\supset\SU(3)\otimes\U(1)$ the $\SU(3)\otimes\U(1)$ irrep $\mu$ in
the l.h.s. is represented as $(\nu,C)$, where $\nu$ is the $\SU(3)$ irrep and
$C$ is the charm quantum number. Now $\Sigma$ stands for the
$(\SU(4);\SU(3)\otimes\U(1))$ irrep $(\mathbf{20};\mathbf{8},0)$ while $\pi$
stands for $(\mathbf{15};\mathbf{8},0)$. ($\rho$ is not used here. It falls in
the same representation $\mathbf{15}$ of $\SU(4)$ as $\pi$ and we choose to
use the pseudoscalars to label the states.)

Finally, for $\SU(3)\supset\SU(2)\otimes\U(1)$ the $\SU(3)\otimes\U(1)$ irrep
$\mu$ in the l.h.s. is represented as $(I,Y)$, where $I$ is the isospin and
$Y$ the hypercharge, and each particle symbol labels a complete
isospin-hypercharge multiplet.

\section{Method of construction and phase conventions}
\label{sec:4}

The fundamental and antifundamental representations of $\SU(n)$ can be
realized by the ladder operators:
\begin{eqnarray}
E^i_j|k\rangle &=& \delta^i_k|j\rangle-\frac{1}{n}\delta^i_j|k\rangle,
\nonumber \\ E^i_j|\bar{k}\rangle &=&
-\delta^k_j|\bar{i}\rangle+\frac{1}{n}\delta^i_j|\bar{k}\rangle, \quad
i,j,k,=1,\ldots, n ,
\label{eq:19}
\end{eqnarray}
which fulfill the su($n$) algebra relations
\begin{equation}
[E^i_j,E^k_l] = \delta^i_l E^k_j-\delta^k_j E^i_l
,
\quad
(E^i_j)^\dagger=E^j_i
\,.
\end{equation}
These operators act the on a tensor product representation in the usual way
\begin{equation}
E^i_j(|A\rangle\otimes|B\rangle)
=
(E^i_j|A\rangle)\otimes|B\rangle
+
|A\rangle\otimes(E^i_j|B\rangle)
.
\end{equation}

For $\SU(8)$, using the tensor products $\mathbf{8^*}\otimes\mathbf{8}$ and
$\mathbf{8}\otimes\mathbf{8}\otimes\mathbf{8}$ we construct the adjoint,
$\mathbf{63}$, and symmetric, $\mathbf{120}$, representations, corresponding
to mesons and baryons, respectively. The CG series of
$\mathbf{63}\otimes\mathbf{63}$ and $\mathbf{120}\otimes\mathbf{63}$ are
resolved by the standard method of extracting the state with highest weight
and applying ladder operators on it to fill a full highest weight
representation. The method is then repeated recursively for the orthogonal
spaces. The same technique is applied for the other groups, SU(6), SU(4) and
SU(3). As usual higher dimension always implies higher weight. The
representations are ordered as in table \ref{tab:3}.

In the CG series considered the degeneracy label $\sigma$ is often redundant.
When needed the symmetric combination, label $s$, is taken to be of highest
weight than the antisymmetric one, label $a$.

\subsection{Flavor groups}

To fix the phases within an irrep of a flavor group, $\SU(4)$ or $\SU(3)$, we
adopt the prescription of \cite{Baird:1964zm}, namely, we demand that the
matrix elements between basis states of the ladder operators of form
$E^i_{i+1}$ should be nonnegative. This produces a standard basis, by
definition.\footnote{In this sense, the basis $|i\rangle$ in Eq.~(\ref{eq:19})
is standard, whereas $|\bar{i}\rangle$ is not.} Identifying, as usual, the states $|i\rangle$, $i=1,2,3,4$ with
$u$, $d$, $s$, $c$ (in this order) this prescription differs form the $\SU(3)$
choice in \cite{deSwart:1963gc}.

To fix the relative phases between the irreps in the CG series of $\SU(4)$, we
fix the sign of the state with highest weight, i.e., corresponding to
the highest SU(3) irrep. We do this is the usual way, namely, we order the
uncoupled states attending first to $\mu_1$, if necessary to $\mu_2$, and
finally to $\gamma^\prime$. The sign of the highest coupled state is then
chosen so that it has a positive overlap with the highest uncoupled
state. E.g., in $\mathbf{15}\otimes\mathbf{15}$ to give $\mathbf{15_s}$, the
coupled state with highest weight is $|\mathbf{15_s};\mathbf{8},0\rangle$. The
highest uncoupled state is $|(\pi,\pi)_s\rangle$.  Everything is similar for
$\SU(3)$. The prescription adopted is just the natural extension of that in
$\SU(2)$.

\subsection{Spin-flavor groups}

For the group $\SU(8)$ we are interested in its reduction under
$\SU(4)\otimes\SU(2)$. Therefore the prescription in \cite{Baird:1964zm},
devised for $\SU(n)\supset\SU(n-1)\otimes\U(1)$, does not directly apply. We
do apply \cite{Baird:1964zm} for the relative phases between states inside
each $\SU(4)\otimes\SU(2)$ irrep, but the phase between two such irreps in the
reduction of an irrep of $\SU(8)$ is to be fixed. Also, because the reduction
$\SU(8)\supset\SU(4)\otimes\SU(2)$ is not a canonical one, an
$\SU(4)\otimes\SU(2)$ irrep can appear several times (label $\gamma$ in
Eq.~(\ref{eq:7})) and this has also to be fixed.  In addition, the phase of
each $\SU(8)$ irrep in the CG series is to be settled.

There is no widely accepted way of defining standard bases of the irreps of
$\SU(8)\supset\SU(4)\otimes\SU(2)$. Rather than introducing such general
prescription we adopt some concrete choices in what follows. Everything we
say here for $\SU(8)$ can be immediately translated to $\SU(6)$.  

Let us consider the irrep $\mathbf{63}$ obtained from
$\mathbf{8^*}\otimes\mathbf{8}$ and the irrep $\mathbf{120}$ obtained from
$\mathbf{8}\otimes\mathbf{8}\otimes\mathbf{8}$.  Their reduction under
$\SU(4)\otimes\SU(2)$ can be looked up in table \ref{tab:2}.  For them we
choose
\begin{eqnarray}
\langle\pi;1,0|E^{u,+}_{u,-}|\rho;1,1\rangle > 0 ,
\nonumber \\
\langle\omega_1;0,1|E^{u,+}_{d,+}|\rho;1,1\rangle > 0 ,
\nonumber \\
\langle\Sigma;1,\frac{1}{2}|E^{u,+}_{u,-}|\Sigma^*;1,\frac{3}{2}\rangle > 0 .
\label{eq:21}
\end{eqnarray}
Here we have used the notation $|p,I_z,J_z\rangle$ where $p$ denotes a
concrete state $R,\mu,\nu,I,J,Y,C$ (see table \ref{tab:3}). Also
$(u,+),(d,+),\ldots,(c,-)$ are the 8 labels $i$ in the ladder operators
$E^i_j$ of $\SU(8)$. The two first relations fix the phases between
$\mathbf{15_1}$, $\mathbf{1_3}$ and $\mathbf{15_3}$ in $\mathbf{63}$, whereas
the second relation fixes the phase between $\mathbf{20_2}$ and
$\mathbf{20^\prime_4}$ in $\mathbf{120}$.

For all the $\SU(4)\otimes\SU(2)$ irreps produced through the products
$\mathbf{63}\otimes\mathbf{63}$ and $\mathbf{120}\otimes\mathbf{63}$, we adopt
a common procedure which is customary and extends that used to resolve the CG
series in $\SU(n)\supset\SU(n-1)\otimes\U(1)$.  This is as follows. The
uncoupled states (r.h.s. in Eq.~(\ref{eq:7}) or in Eq.~(\ref{eq:11})) are
given a well defined order. Then the first coupled state is taken to have the
maximum overlap with the first uncoupled state. (If this overlap is zero the
next uncoupled state is taken instead, etc.)  Next, the second coupled state
is chosen as the state orthogonal to the first one having the maximum overlap
with the second uncoupled state. And so on, recursively.

So, for instance, in
$|\mathbf{63}\otimes\mathbf{63};\mathbf{63_s};\mathbf{15_3}\rangle$,
$|(\rho,\rho)_a\rangle$ is the highest uncoupled state ($\rho$ being of higher
weight than $\pi$ and $\omega_1$) and so its coefficient is positive. In this
example the label $\gamma$ was not needed. A more complicated case is
$|\mathbf{120}\otimes\mathbf{63};\mathbf{4752};\mathbf{20_{\gamma,4}}\rangle$
with $\gamma=s,a,b$. The relevant uncoupled states are, listed from highest to
lowest, $|\Delta,\rho\rangle$, $|\Delta,\pi\rangle$,
$|(\Sigma,\rho)_s\rangle$, $|(\Sigma,\rho)_a\rangle$, and
$|\Sigma,\omega_1\rangle$. Hence, $|\mathbf{20_{s,4}}\rangle$ has positive
overlap with $|\Delta,\rho\rangle$, $|\mathbf{20_{a,4}}\rangle$ has no overlap
with $|\Delta,\rho\rangle$ and positive overlap with $|\Delta,\pi\rangle$, and
$|\mathbf{20_{b,4}}\rangle$ has no overlap with $|\Delta,\rho\rangle$ or
$|\Delta,\pi\rangle$ and positive overlap with $|(\Sigma,\rho)_s\rangle$.

We point out that this prescription, although simple enough, does not
automatically produce standard bases. The bases obtained for a given $\SU(8)$
irrep will depend on how that irrep is obtained. For instance, for the
$\mathbf{63}$ obtained from $\mathbf{8^*}\otimes\mathbf{8}$ we have fixed the
relative phases between $\mathbf{15_1}$, $\mathbf{15_3}$ and $\mathbf{1_3}$ as
in Eq.~(\ref{eq:21}). However, the product $\mathbf{63}\otimes\mathbf{63}$
produces two $\mathbf{63}$ irreps (namely, $\mathbf{63_s}$ and
$\mathbf{63_a}$). For them the relative phases between $\mathbf{15_1}$,
$\mathbf{15_3}$ and $\mathbf{1_3}$ are fixed instead from the ``coupling
method'' just described.  Unfortunately, it turns out to violate the two
inequalities in (\ref{eq:21}). And the same violation takes places for the new
irrep $\mathbf{120}$ generated from $\mathbf{120}\otimes\mathbf{63}$ for the
third inequality. Of course, we could just redefine the necessary signs
within these $\mathbf{63_s}$, $\mathbf{63_a}$ and $\mathbf{120}$, but we
prefer to be systematic rather than to enforce standard bases for particular
irreps of $\SU(8)$.

Regarding the order of the uncoupled states, there is a subtlety. In general,
the order depends first on $\mu_1$, then on $\mu_2$ and then on
$\gamma^\prime$.  However, in $\mathbf{63}\otimes\mathbf{120}$, we attend
first to $\mu_2$ (the baryon state) and then to $\mu_1$ (the meson state).
This is necessary to have a well defined $\xi$ in Eq.~(\ref{eq:18}) in all
cases. For instance, in
$|\mathbf{63}\otimes\mathbf{120};\mathbf{4752};\mathbf{20_{\gamma,4}}\rangle$
we want $|\pi,\Delta\rangle$ to be of higher weight than
$|(\rho,\Sigma)_s\rangle$, so that under exchange of labels $1$ and $2$
(Eq.~(\ref{eq:18})), 
$|\mathbf{63}\otimes\mathbf{120};\mathbf{4752};\mathbf{20_{a,4}}\rangle$
maps to
$|\mathbf{120}\otimes\mathbf{63};\mathbf{4752};\mathbf{20_{a,4}}\rangle$
and
$|\mathbf{63}\otimes\mathbf{120};\mathbf{4752};\mathbf{20_{b,4}}\rangle$
maps to
$|\mathbf{120}\otimes\mathbf{63};\mathbf{4752};\mathbf{20_{b,4}}\rangle$.

\begin{acknowledgments}
We thank J. Nieves for discussions. Research supported by DGI under
contract FIS2008-01143, Junta de Andaluc\'{\i}a grant FQM-225, the Spanish
Consolider-Ingenio 2010 Programme CPAN contract CSD2007-00042, and the European
Community-Research Infrastructure Integrating Activity {\em Study of Strongly
Interacting Matter} (HadronPhysics2, Grant Agreement no. 227431)
under the 7th Framework Programme of EU.
\end{acknowledgments}

\clearpage
\newpage

\appendix

\begin{widetext}
\section{Tables of scalar factors of SU(8)}

\begin{eqnarray}
|\rho\rangle &=& |\mathbf{63};\mathbf{15_3}\rangle
,
\quad
|\pi\rangle = |\mathbf{63};\mathbf{15_1}\rangle
,
\quad
|\omega_1\rangle = |\mathbf{63};\mathbf{1_3}\rangle
,
\nonumber\\
|\Delta\rangle &=& |\mathbf{120};\mathbf{20^\prime_4}\rangle
,
\quad
|\Sigma\rangle = |\mathbf{120};\mathbf{20_2}\rangle
.
\nonumber
\end{eqnarray}

\subsection{SU(8): $ \mathbf{63} \otimes \mathbf{63}$} \par
\begin{eqnarray}
(+)~~&
|\mathbf{1};\mathbf{1_1}\rangle
&=
\phantom{+}\sqrt{\frac{5}{7}}|\rho,\rho\rangle
-\sqrt{\frac{1}{21}}|\omega_1,\omega_1\rangle
-\sqrt{\frac{5}{21}}|\pi,\pi\rangle
\nonumber \\
(+)~~&
|\mathbf{720};\mathbf{1_1}\rangle
&=
\phantom{+}\sqrt{\frac{1}{28}}|\rho,\rho\rangle
-\sqrt{\frac{15}{28}}|\omega_1,\omega_1\rangle
+\sqrt{\frac{3}{7}}|\pi,\pi\rangle
\nonumber \\
(+)~~&
|\mathbf{1232};\mathbf{1_1}\rangle
&=
\phantom{+}\sqrt{\frac{1}{4}}|\rho,\rho\rangle
+\sqrt{\frac{5}{12}}|\omega_1,\omega_1\rangle
+\sqrt{\frac{1}{3}}|\pi,\pi\rangle
\end{eqnarray}
\begin{eqnarray}
(+)~~&
|\mathbf{63_s};\mathbf{1_3}\rangle
&=
\phantom{+}|\rho,\pi\rangle_S
\nonumber \\
(-)~~&
|\mathbf{63_a};\mathbf{1_3}\rangle
&=
\phantom{+}\sqrt{\frac{15}{16}}|\rho,\rho\rangle
-\sqrt{\frac{1}{16}}|\omega_1,\omega_1\rangle
\nonumber \\
(-)~~&
|\mathbf{945};\mathbf{1_3}\rangle
&=
\phantom{+}\sqrt{\frac{1}{32}}|\rho,\rho\rangle
+\sqrt{\frac{1}{2}}|\rho,\pi\rangle_A
+\sqrt{\frac{15}{32}}|\omega_1,\omega_1\rangle
\nonumber \\
(-)~~&
|\mathbf{945^*};\mathbf{1_3}\rangle
&=
\phantom{+}\sqrt{\frac{1}{32}}|\rho,\rho\rangle
-\sqrt{\frac{1}{2}}|\rho,\pi\rangle_A
+\sqrt{\frac{15}{32}}|\omega_1,\omega_1\rangle
\end{eqnarray}
\begin{eqnarray}
(+)~~&
|\mathbf{720};\mathbf{1_5}\rangle
&=
\phantom{+}\sqrt{\frac{5}{8}}|\rho,\rho\rangle
+\sqrt{\frac{3}{8}}|\omega_1,\omega_1\rangle
\nonumber \\
(+)~~&
|\mathbf{1232};\mathbf{1_5}\rangle
&=
\phantom{+}\sqrt{\frac{3}{8}}|\rho,\rho\rangle
-\sqrt{\frac{5}{8}}|\omega_1,\omega_1\rangle
\end{eqnarray}
\begin{eqnarray}
(+)~~&
|\mathbf{63_s};\mathbf{15_1}\rangle
&=
\phantom{+}\sqrt{\frac{3}{5}}|(\rho,\rho)_s\rangle
+\sqrt{\frac{1}{5}}|\rho,\omega_1\rangle_S
-\sqrt{\frac{1}{5}}|(\pi,\pi)_s\rangle
\nonumber \\
(-)~~&
|\mathbf{63_a};\mathbf{15_1}\rangle
&=
\phantom{+}\sqrt{\frac{3}{4}}|(\rho,\rho)_a\rangle
-\sqrt{\frac{1}{4}}|(\pi,\pi)_a\rangle
\nonumber \\
(+)~~&
|\mathbf{720};\mathbf{15_1}\rangle
&=
\phantom{+}\sqrt{\frac{1}{2}}|\rho,\omega_1\rangle_S
+\sqrt{\frac{1}{2}}|(\pi,\pi)_s\rangle
\nonumber \\
(-)~~&
|\mathbf{945};\mathbf{15_1}\rangle
&=
\phantom{+}\sqrt{\frac{1}{8}}|(\rho,\rho)_a\rangle
+\sqrt{\frac{1}{2}}|\rho,\omega_1\rangle_A
+\sqrt{\frac{3}{8}}|(\pi,\pi)_a\rangle
\nonumber \\
(-)~~&
|\mathbf{945^*};\mathbf{15_1}\rangle
&=
\phantom{+}\sqrt{\frac{1}{8}}|(\rho,\rho)_a\rangle
-\sqrt{\frac{1}{2}}|\rho,\omega_1\rangle_A
+\sqrt{\frac{3}{8}}|(\pi,\pi)_a\rangle
\nonumber \\
(+)~~&
|\mathbf{1232};\mathbf{15_1}\rangle
&=
\phantom{+}\sqrt{\frac{2}{5}}|(\rho,\rho)_s\rangle
-\sqrt{\frac{3}{10}}|\rho,\omega_1\rangle_S
+\sqrt{\frac{3}{10}}|(\pi,\pi)_s\rangle
\end{eqnarray}
\begin{eqnarray}
(+)~~&
|\mathbf{63_s};\mathbf{15_3}\rangle
&=
\phantom{+}\sqrt{\frac{8}{15}}|(\rho,\rho)_a\rangle
+\sqrt{\frac{2}{5}}|(\rho,\pi)_s\rangle_S
+\sqrt{\frac{1}{15}}|\omega_1,\pi\rangle_S
\nonumber \\
(-)~~&
|\mathbf{63_a};\mathbf{15_3}\rangle
&=
\phantom{+}\sqrt{\frac{3}{8}}|(\rho,\rho)_s\rangle
+\sqrt{\frac{1}{8}}|\rho,\omega_1\rangle_S
+\sqrt{\frac{1}{2}}|(\rho,\pi)_a\rangle_S
\nonumber \\
(+)~~&
|\mathbf{720};\mathbf{15_{s,3}}\rangle
&=
\phantom{+}\sqrt{\frac{1}{6}}|(\rho,\rho)_a\rangle
-\sqrt{\frac{1}{2}}|(\rho,\pi)_s\rangle_S
+\sqrt{\frac{1}{3}}|\omega_1,\pi\rangle_S
\nonumber \\
(+)~~&
|\mathbf{720};\mathbf{15_{a,3}}\rangle
&=
\phantom{+}\sqrt{\frac{1}{2}}|\rho,\omega_1\rangle_A
-\sqrt{\frac{1}{2}}|(\rho,\pi)_a\rangle_A
\nonumber \\
(-)~~&
|\mathbf{945};\mathbf{15_{s,3}}\rangle
&=
\phantom{+}\sqrt{\frac{5}{16}}|(\rho,\rho)_s\rangle
-\sqrt{\frac{3}{80}}|\rho,\omega_1\rangle_S
+\sqrt{\frac{1}{5}}|(\rho,\pi)_s\rangle_A
-\sqrt{\frac{3}{20}}|(\rho,\pi)_a\rangle_S
\nonumber \\ &&
-\sqrt{\frac{3}{10}}|\omega_1,\pi\rangle_A
\nonumber \\
(-)~~&
|\mathbf{945};\mathbf{15_{a,3}}\rangle
&=
\phantom{+}\sqrt{\frac{2}{5}}|\rho,\omega_1\rangle_S
-\sqrt{\frac{3}{10}}|(\rho,\pi)_s\rangle_A
-\sqrt{\frac{1}{10}}|(\rho,\pi)_a\rangle_S
-\sqrt{\frac{1}{5}}|\omega_1,\pi\rangle_A
\nonumber \\
(-)~~&
|\mathbf{945^*};\mathbf{15_{s,3}}\rangle
&=
\phantom{+}\sqrt{\frac{5}{16}}|(\rho,\rho)_s\rangle
-\sqrt{\frac{3}{80}}|\rho,\omega_1\rangle_S
-\sqrt{\frac{1}{5}}|(\rho,\pi)_s\rangle_A
-\sqrt{\frac{3}{20}}|(\rho,\pi)_a\rangle_S
\nonumber \\ &&
+\sqrt{\frac{3}{10}}|\omega_1,\pi\rangle_A
\nonumber \\
(-)~~&
|\mathbf{945^*};\mathbf{15_{a,3}}\rangle
&=
\phantom{+}\sqrt{\frac{2}{5}}|\rho,\omega_1\rangle_S
+\sqrt{\frac{3}{10}}|(\rho,\pi)_s\rangle_A
-\sqrt{\frac{1}{10}}|(\rho,\pi)_a\rangle_S
+\sqrt{\frac{1}{5}}|\omega_1,\pi\rangle_A
\nonumber \\
(+)~~&
|\mathbf{1232};\mathbf{15_{s,3}}\rangle
&=
\phantom{+}\sqrt{\frac{3}{10}}|(\rho,\rho)_a\rangle
-\sqrt{\frac{1}{10}}|(\rho,\pi)_s\rangle_S
-\sqrt{\frac{3}{5}}|\omega_1,\pi\rangle_S
\nonumber \\
(+)~~&
|\mathbf{1232};\mathbf{15_{a,3}}\rangle
&=
\phantom{+}\sqrt{\frac{1}{2}}|\rho,\omega_1\rangle_A
+\sqrt{\frac{1}{2}}|(\rho,\pi)_a\rangle_A
\end{eqnarray}
\begin{eqnarray}
(+)~~&
|\mathbf{720};\mathbf{15_5}\rangle
&=
\phantom{+}\sqrt{\frac{3}{4}}|(\rho,\rho)_s\rangle
-\sqrt{\frac{1}{4}}|\rho,\omega_1\rangle_S
\nonumber \\
(-)~~&
|\mathbf{945};\mathbf{15_5}\rangle
&=
\phantom{+}\sqrt{\frac{1}{2}}|(\rho,\rho)_a\rangle
-\sqrt{\frac{1}{2}}|\rho,\omega_1\rangle_A
\nonumber \\
(-)~~&
|\mathbf{945^*};\mathbf{15_5}\rangle
&=
\phantom{+}\sqrt{\frac{1}{2}}|(\rho,\rho)_a\rangle
+\sqrt{\frac{1}{2}}|\rho,\omega_1\rangle_A
\nonumber \\
(+)~~&
|\mathbf{1232};\mathbf{15_5}\rangle
&=
\phantom{+}\sqrt{\frac{1}{4}}|(\rho,\rho)_s\rangle
+\sqrt{\frac{3}{4}}|\rho,\omega_1\rangle_S
\end{eqnarray}
\begin{eqnarray}
(+)~~&
|\mathbf{720};\mathbf{20^{\prime\prime}_1}\rangle
&=
\phantom{+}\sqrt{\frac{1}{4}}|\rho,\rho\rangle
-\sqrt{\frac{3}{4}}|\pi,\pi\rangle
\nonumber \\
(+)~~&
|\mathbf{1232};\mathbf{20^{\prime\prime}_1}\rangle
&=
\phantom{+}\sqrt{\frac{3}{4}}|\rho,\rho\rangle
+\sqrt{\frac{1}{4}}|\pi,\pi\rangle
\end{eqnarray}
\begin{eqnarray}
(+)~~&
|\mathbf{720};\mathbf{20^{\prime\prime}_3}\rangle
&=
\phantom{+}|\rho,\pi\rangle_S
\nonumber \\
(-)~~&
|\mathbf{945};\mathbf{20^{\prime\prime}_3}\rangle
&=
\phantom{+}\sqrt{\frac{1}{2}}|\rho,\rho\rangle
+\sqrt{\frac{1}{2}}|\rho,\pi\rangle_A
\nonumber \\
(-)~~&
|\mathbf{945^*};\mathbf{20^{\prime\prime}_3}\rangle
&=
\phantom{+}\sqrt{\frac{1}{2}}|\rho,\rho\rangle
-\sqrt{\frac{1}{2}}|\rho,\pi\rangle_A
\end{eqnarray}
\begin{eqnarray}
(+)~~&
|\mathbf{720};\mathbf{20^{\prime\prime}_5}\rangle
&=
\phantom{+}|\rho,\rho\rangle
\end{eqnarray}
\begin{eqnarray}
(-)~~&
|\mathbf{945};\mathbf{45_1}\rangle
&=
\phantom{+}\sqrt{\frac{1}{4}}|\rho,\rho\rangle
-\sqrt{\frac{3}{4}}|\pi,\pi\rangle
\nonumber \\
(-)~~&
|\mathbf{945^*};\mathbf{45_1}\rangle
&=
\phantom{+}\sqrt{\frac{3}{4}}|\rho,\rho\rangle
+\sqrt{\frac{1}{4}}|\pi,\pi\rangle
\end{eqnarray}
\begin{eqnarray}
(+)~~&
|\mathbf{720};\mathbf{45_3}\rangle
&=
\phantom{+}\sqrt{\frac{1}{2}}|\rho,\rho\rangle
+\sqrt{\frac{1}{2}}|\rho,\pi\rangle_A
\nonumber \\
(-)~~&
|\mathbf{945};\mathbf{45_3}\rangle
&=
\phantom{+}|\rho,\pi\rangle_S
\nonumber \\
(+)~~&
|\mathbf{1232};\mathbf{45_3}\rangle
&=
\phantom{+}\sqrt{\frac{1}{2}}|\rho,\rho\rangle
-\sqrt{\frac{1}{2}}|\rho,\pi\rangle_A
\end{eqnarray}
\begin{eqnarray}
(-)~~&
|\mathbf{945};\mathbf{45_5}\rangle
&=
\phantom{+}|\rho,\rho\rangle
\end{eqnarray}
\begin{eqnarray}
(-)~~&
|\mathbf{945};\mathbf{45^*_1}\rangle
&=
\phantom{+}\sqrt{\frac{3}{4}}|\rho,\rho\rangle
+\sqrt{\frac{1}{4}}|\pi,\pi\rangle
\nonumber \\
(-)~~&
|\mathbf{945^*};\mathbf{45^*_1}\rangle
&=
\phantom{+}\sqrt{\frac{1}{4}}|\rho,\rho\rangle
-\sqrt{\frac{3}{4}}|\pi,\pi\rangle
\end{eqnarray}
\begin{eqnarray}
(+)~~&
|\mathbf{720};\mathbf{45^*_3}\rangle
&=
\phantom{+}\sqrt{\frac{1}{2}}|\rho,\rho\rangle
-\sqrt{\frac{1}{2}}|\rho,\pi\rangle_A
\nonumber \\
(-)~~&
|\mathbf{945^*};\mathbf{45^*_3}\rangle
&=
\phantom{+}|\rho,\pi\rangle_S
\nonumber \\
(+)~~&
|\mathbf{1232};\mathbf{45^*_3}\rangle
&=
\phantom{+}\sqrt{\frac{1}{2}}|\rho,\rho\rangle
+\sqrt{\frac{1}{2}}|\rho,\pi\rangle_A
\end{eqnarray}
\begin{eqnarray}
(-)~~&
|\mathbf{945^*};\mathbf{45^*_5}\rangle
&=
\phantom{+}|\rho,\rho\rangle
\end{eqnarray}
\begin{eqnarray}
(+)~~&
|\mathbf{720};\mathbf{84_1}\rangle
&=
\phantom{+}\sqrt{\frac{3}{4}}|\rho,\rho\rangle
+\sqrt{\frac{1}{4}}|\pi,\pi\rangle
\nonumber \\
(+)~~&
|\mathbf{1232};\mathbf{84_1}\rangle
&=
\phantom{+}\sqrt{\frac{1}{4}}|\rho,\rho\rangle
-\sqrt{\frac{3}{4}}|\pi,\pi\rangle
\end{eqnarray}
\begin{eqnarray}
(-)~~&
|\mathbf{945};\mathbf{84_3}\rangle
&=
\phantom{+}\sqrt{\frac{1}{2}}|\rho,\rho\rangle
-\sqrt{\frac{1}{2}}|\rho,\pi\rangle_A
\nonumber \\
(-)~~&
|\mathbf{945^*};\mathbf{84_3}\rangle
&=
\phantom{+}\sqrt{\frac{1}{2}}|\rho,\rho\rangle
+\sqrt{\frac{1}{2}}|\rho,\pi\rangle_A
\nonumber \\
(+)~~&
|\mathbf{1232};\mathbf{84_3}\rangle
&=
\phantom{+}|\rho,\pi\rangle_S
\end{eqnarray}
\begin{eqnarray}
(+)~~&
|\mathbf{1232};\mathbf{84_5}\rangle
&=
\phantom{+}|\rho,\rho\rangle
\end{eqnarray}
\subsection{SU(8): $ \mathbf{120} \otimes \mathbf{63}$} \par
\begin{eqnarray}
(+)~~&
|\mathbf{168};\mathbf{4^*_2}\rangle
&=
\phantom{+}\sqrt{\frac{3}{4}}|\Sigma,\rho\rangle
-\sqrt{\frac{1}{4}}|\Sigma,\pi\rangle
\nonumber \\
(+)~~&
|\mathbf{4752};\mathbf{4^*_2}\rangle
&=
\phantom{+}\sqrt{\frac{1}{4}}|\Sigma,\rho\rangle
+\sqrt{\frac{3}{4}}|\Sigma,\pi\rangle
\end{eqnarray}
\begin{eqnarray}
(-)~~&
|\mathbf{4752};\mathbf{4^*_4}\rangle
&=
\phantom{+}|\Sigma,\rho\rangle
\end{eqnarray}
\begin{eqnarray}
(+)~~&
|\mathbf{120};\mathbf{20_2}\rangle
&=
\phantom{+}\sqrt{\frac{32}{77}}|\Delta,\rho\rangle
+\sqrt{\frac{256}{1001}}|(\Sigma,\rho)_s\rangle
-\sqrt{\frac{21}{143}}|(\Sigma,\rho)_a\rangle
+\sqrt{\frac{1}{77}}|\Sigma,\omega_1\rangle
\nonumber \\ &&
+\sqrt{\frac{13}{77}}|(\Sigma,\pi)_a\rangle
\nonumber \\
(+)~~&
|\mathbf{168};\mathbf{20_2}\rangle
&=
\phantom{+}\sqrt{\frac{2}{5}}|\Delta,\rho\rangle
-\sqrt{\frac{4}{65}}|(\Sigma,\rho)_s\rangle
+\sqrt{\frac{15}{52}}|(\Sigma,\rho)_a\rangle
-\sqrt{\frac{1}{20}}|\Sigma,\omega_1\rangle
\nonumber \\ &&
+\sqrt{\frac{12}{65}}|(\Sigma,\pi)_s\rangle
-\sqrt{\frac{1}{65}}|(\Sigma,\pi)_a\rangle
\nonumber \\
(+)~~&
|\mathbf{2520};\mathbf{20_2}\rangle
&=
\phantom{+}\sqrt{\frac{2}{33}}|\Delta,\rho\rangle
-\sqrt{\frac{625}{1716}}|(\Sigma,\rho)_s\rangle
+\sqrt{\frac{1}{143}}|(\Sigma,\rho)_a\rangle
+\sqrt{\frac{3}{11}}|\Sigma,\omega_1\rangle
\nonumber \\ &&
-\sqrt{\frac{11}{52}}|(\Sigma,\pi)_s\rangle
+\sqrt{\frac{12}{143}}|(\Sigma,\pi)_a\rangle
\nonumber \\
(+)~~&
|\mathbf{4752};\mathbf{20_{s,2}}\rangle
&=
\phantom{+}\sqrt{\frac{13}{105}}|\Delta,\rho\rangle
-\sqrt{\frac{121}{35490}}|(\Sigma,\rho)_s\rangle
-\sqrt{\frac{35}{338}}|(\Sigma,\rho)_a\rangle
-\sqrt{\frac{27}{910}}|\Sigma,\omega_1\rangle
\nonumber \\ &&
-\sqrt{\frac{343}{1690}}|(\Sigma,\pi)_s\rangle
-\sqrt{\frac{3174}{5915}}|(\Sigma,\pi)_a\rangle
\nonumber \\
(-)~~&
|\mathbf{4752};\mathbf{20_{a,2}}\rangle
&=
\phantom{+}\sqrt{\frac{213}{676}}|(\Sigma,\rho)_s\rangle
+\sqrt{\frac{4900}{11999}}|(\Sigma,\rho)_a\rangle
+\sqrt{\frac{108}{923}}|\Sigma,\omega_1\rangle
-\sqrt{\frac{5929}{47996}}|(\Sigma,\pi)_s\rangle
\nonumber \\ &&
-\sqrt{\frac{432}{11999}}|(\Sigma,\pi)_a\rangle
\nonumber \\
(+)~~&
|\mathbf{4752};\mathbf{20_{b,2}}\rangle
&=
\phantom{+}\sqrt{\frac{13}{284}}|(\Sigma,\rho)_a\rangle
-\sqrt{\frac{147}{284}}|\Sigma,\omega_1\rangle
-\sqrt{\frac{256}{923}}|(\Sigma,\pi)_s\rangle
+\sqrt{\frac{147}{923}}|(\Sigma,\pi)_a\rangle
\end{eqnarray}
\begin{eqnarray}
(-)~~&
|\mathbf{168};\mathbf{20_4}\rangle
&=
\phantom{+}\sqrt{\frac{1}{2}}|\Delta,\rho\rangle
-\sqrt{\frac{3}{10}}|\Delta,\pi\rangle
+\sqrt{\frac{8}{65}}|(\Sigma,\rho)_s\rangle
+\sqrt{\frac{27}{520}}|(\Sigma,\rho)_a\rangle
\nonumber \\ &&
-\sqrt{\frac{1}{40}}|\Sigma,\omega_1\rangle
\nonumber \\
(-)~~&
|\mathbf{2520};\mathbf{20_4}\rangle
&=
\phantom{+}\sqrt{\frac{5}{24}}|\Delta,\rho\rangle
+\sqrt{\frac{1}{8}}|\Delta,\pi\rangle
+\sqrt{\frac{2}{39}}|(\Sigma,\rho)_s\rangle
-\sqrt{\frac{25}{104}}|(\Sigma,\rho)_a\rangle
\nonumber \\ &&
+\sqrt{\frac{3}{8}}|\Sigma,\omega_1\rangle
\nonumber \\
(-)~~&
|\mathbf{4752};\mathbf{20_{s,4}}\rangle
&=
\phantom{+}\sqrt{\frac{7}{24}}|\Delta,\rho\rangle
+\sqrt{\frac{7}{40}}|\Delta,\pi\rangle
-\sqrt{\frac{578}{1365}}|(\Sigma,\rho)_s\rangle
+\sqrt{\frac{7}{520}}|(\Sigma,\rho)_a\rangle
\nonumber \\ &&
-\sqrt{\frac{27}{280}}|\Sigma,\omega_1\rangle
\nonumber \\
(+)~~&
|\mathbf{4752};\mathbf{20_{a,4}}\rangle
&=
\phantom{+}\sqrt{\frac{2}{5}}|\Delta,\pi\rangle
+\sqrt{\frac{24}{65}}|(\Sigma,\rho)_s\rangle
+\sqrt{\frac{81}{520}}|(\Sigma,\rho)_a\rangle
-\sqrt{\frac{3}{40}}|\Sigma,\omega_1\rangle
\nonumber \\
(+)~~&
|\mathbf{4752};\mathbf{20_{b,4}}\rangle
&=
\phantom{+}\sqrt{\frac{3}{91}}|(\Sigma,\rho)_s\rangle
-\sqrt{\frac{7}{13}}|(\Sigma,\rho)_a\rangle
-\sqrt{\frac{3}{7}}|\Sigma,\omega_1\rangle
\end{eqnarray}
\begin{eqnarray}
(+)~~&
|\mathbf{4752};\mathbf{20_6}\rangle
&=
\phantom{+}|\Delta,\rho\rangle
\end{eqnarray}
\begin{eqnarray}
(-)~~&
|\mathbf{168};\mathbf{20^\prime_2}\rangle
&=
\phantom{+}\sqrt{\frac{7}{10}}|\Delta,\rho\rangle
-\sqrt{\frac{1}{10}}|\Delta,\omega_1\rangle
+\sqrt{\frac{1}{20}}|\Sigma,\rho\rangle
+\sqrt{\frac{3}{20}}|\Sigma,\pi\rangle
\nonumber \\
(-)~~&
|\mathbf{2520};\mathbf{20^\prime_2}\rangle
&=
\phantom{+}\sqrt{\frac{7}{48}}|\Delta,\rho\rangle
+\sqrt{\frac{3}{16}}|\Delta,\omega_1\rangle
+\sqrt{\frac{1}{6}}|\Sigma,\rho\rangle
-\sqrt{\frac{1}{2}}|\Sigma,\pi\rangle
\nonumber \\
(-)~~&
|\mathbf{4752};\mathbf{20^\prime_{s,2}}\rangle
&=
\phantom{+}\sqrt{\frac{37}{240}}|\Delta,\rho\rangle
+\sqrt{\frac{189}{2960}}|\Delta,\omega_1\rangle
-\sqrt{\frac{847}{1110}}|\Sigma,\rho\rangle
-\sqrt{\frac{7}{370}}|\Sigma,\pi\rangle
\nonumber \\
(+)~~&
|\mathbf{4752};\mathbf{20^\prime_{a,2}}\rangle
&=
\phantom{+}\sqrt{\frac{24}{37}}|\Delta,\omega_1\rangle
+\sqrt{\frac{3}{148}}|\Sigma,\rho\rangle
+\sqrt{\frac{49}{148}}|\Sigma,\pi\rangle
\end{eqnarray}
\begin{eqnarray}
(+)~~&
|\mathbf{120};\mathbf{20^\prime_4}\rangle
&=
\phantom{+}\sqrt{\frac{5}{11}}|\Delta,\rho\rangle
-\sqrt{\frac{5}{77}}|\Delta,\omega_1\rangle
-\sqrt{\frac{3}{11}}|\Delta,\pi\rangle
+\sqrt{\frac{16}{77}}|\Sigma,\rho\rangle
\nonumber \\
(+)~~&
|\mathbf{2520};\mathbf{20^\prime_4}\rangle
&=
\phantom{+}\sqrt{\frac{1}{264}}|\Delta,\rho\rangle
+\sqrt{\frac{21}{88}}|\Delta,\omega_1\rangle
+\sqrt{\frac{5}{22}}|\Delta,\pi\rangle
+\sqrt{\frac{35}{66}}|\Sigma,\rho\rangle
\nonumber \\
(+)~~&
|\mathbf{4752};\mathbf{20^\prime_{s,4}}\rangle
&=
\phantom{+}\sqrt{\frac{13}{24}}|\Delta,\rho\rangle
+\sqrt{\frac{27}{728}}|\Delta,\omega_1\rangle
+\sqrt{\frac{5}{26}}|\Delta,\pi\rangle
-\sqrt{\frac{125}{546}}|\Sigma,\rho\rangle
\nonumber \\
(-)~~&
|\mathbf{4752};\mathbf{20^\prime_{a,4}}\rangle
&=
\phantom{+}\sqrt{\frac{60}{91}}|\Delta,\omega_1\rangle
-\sqrt{\frac{4}{13}}|\Delta,\pi\rangle
-\sqrt{\frac{3}{91}}|\Sigma,\rho\rangle
\end{eqnarray}
\begin{eqnarray}
(-)~~&
|\mathbf{2520};\mathbf{20^\prime_6}\rangle
&=
\phantom{+}\sqrt{\frac{9}{16}}|\Delta,\rho\rangle
-\sqrt{\frac{7}{16}}|\Delta,\omega_1\rangle
\nonumber \\
(-)~~&
|\mathbf{4752};\mathbf{20^\prime_6}\rangle
&=
\phantom{+}\sqrt{\frac{7}{16}}|\Delta,\rho\rangle
+\sqrt{\frac{9}{16}}|\Delta,\omega_1\rangle
\end{eqnarray}
\begin{eqnarray}
(-)~~&
|\mathbf{4752};\mathbf{36^*_{s,2}}\rangle
&=
\phantom{+}|\Sigma,\rho\rangle
\nonumber \\
(+)~~&
|\mathbf{4752};\mathbf{36^*_{a,2}}\rangle
&=
\phantom{+}|\Sigma,\pi\rangle
\end{eqnarray}
\begin{eqnarray}
(+)~~&
|\mathbf{4752};\mathbf{36^*_4}\rangle
&=
\phantom{+}|\Sigma,\rho\rangle
\end{eqnarray}
\begin{eqnarray}
(+)~~&
|\mathbf{2520};\mathbf{60^*_2}\rangle
&=
\phantom{+}\sqrt{\frac{3}{4}}|\Sigma,\rho\rangle
+\sqrt{\frac{1}{4}}|\Sigma,\pi\rangle
\nonumber \\
(+)~~&
|\mathbf{4752};\mathbf{60^*_2}\rangle
&=
\phantom{+}\sqrt{\frac{1}{4}}|\Sigma,\rho\rangle
-\sqrt{\frac{3}{4}}|\Sigma,\pi\rangle
\end{eqnarray}
\begin{eqnarray}
(-)~~&
|\mathbf{4752};\mathbf{60^*_4}\rangle
&=
\phantom{+}|\Sigma,\rho\rangle
\end{eqnarray}
\begin{eqnarray}
(+)~~&
|\mathbf{4752};\mathbf{120_2}\rangle
&=
\phantom{+}|\Delta,\rho\rangle
\end{eqnarray}
\begin{eqnarray}
(-)~~&
|\mathbf{2520};\mathbf{120_4}\rangle
&=
\phantom{+}\sqrt{\frac{3}{8}}|\Delta,\rho\rangle
-\sqrt{\frac{5}{8}}|\Delta,\pi\rangle
\nonumber \\
(-)~~&
|\mathbf{4752};\mathbf{120_4}\rangle
&=
\phantom{+}\sqrt{\frac{5}{8}}|\Delta,\rho\rangle
+\sqrt{\frac{3}{8}}|\Delta,\pi\rangle
\end{eqnarray}
\begin{eqnarray}
(+)~~&
|\mathbf{2520};\mathbf{120_6}\rangle
&=
\phantom{+}|\Delta,\rho\rangle
\end{eqnarray}
\begin{eqnarray}
(-)~~&
|\mathbf{2520};\mathbf{140_2}\rangle
&=
\phantom{+}\sqrt{\frac{1}{3}}|\Delta,\rho\rangle
-\sqrt{\frac{1}{6}}|\Sigma,\rho\rangle
+\sqrt{\frac{1}{2}}|\Sigma,\pi\rangle
\nonumber \\
(-)~~&
|\mathbf{4752};\mathbf{140_{s,2}}\rangle
&=
\phantom{+}\sqrt{\frac{2}{3}}|\Delta,\rho\rangle
+\sqrt{\frac{1}{12}}|\Sigma,\rho\rangle
-\sqrt{\frac{1}{4}}|\Sigma,\pi\rangle
\nonumber \\
(-)~~&
|\mathbf{4752};\mathbf{140_{a,2}}\rangle
&=
\phantom{+}\sqrt{\frac{3}{4}}|\Sigma,\rho\rangle
+\sqrt{\frac{1}{4}}|\Sigma,\pi\rangle
\end{eqnarray}
\begin{eqnarray}
(+)~~&
|\mathbf{2520};\mathbf{140_4}\rangle
&=
\phantom{+}\sqrt{\frac{5}{24}}|\Delta,\rho\rangle
+\sqrt{\frac{1}{8}}|\Delta,\pi\rangle
-\sqrt{\frac{2}{3}}|\Sigma,\rho\rangle
\nonumber \\
(+)~~&
|\mathbf{4752};\mathbf{140_{s,4}}\rangle
&=
\phantom{+}\sqrt{\frac{19}{24}}|\Delta,\rho\rangle
-\sqrt{\frac{5}{152}}|\Delta,\pi\rangle
+\sqrt{\frac{10}{57}}|\Sigma,\rho\rangle
\nonumber \\
(-)~~&
|\mathbf{4752};\mathbf{140_{a,4}}\rangle
&=
\phantom{+}\sqrt{\frac{16}{19}}|\Delta,\pi\rangle
+\sqrt{\frac{3}{19}}|\Sigma,\rho\rangle
\end{eqnarray}
\begin{eqnarray}
(-)~~&
|\mathbf{4752};\mathbf{140_6}\rangle
&=
\phantom{+}|\Delta,\rho\rangle
\end{eqnarray}
\section{Tables of scalar factors of SU(6)}
\begin{eqnarray}
|\rho\rangle &=& |\mathbf{35};\mathbf{8_3}\rangle
,
\quad
|\pi\rangle = |\mathbf{35};\mathbf{8_1}\rangle
,
\quad
|\omega_1\rangle = |\mathbf{35};\mathbf{1_3}\rangle
,
\nonumber\\
|\Delta\rangle &=& |\mathbf{56};\mathbf{10_4}\rangle
,
\quad
|\Sigma\rangle = |\mathbf{56};\mathbf{8_2}\rangle
.
\nonumber
\end{eqnarray}

\subsection{SU(6): $ \mathbf{35} \otimes \mathbf{35}$} \par
\begin{eqnarray}
(+)~~&
|\mathbf{1};\mathbf{1_1}\rangle
&=
\phantom{+}\sqrt{\frac{24}{35}}|\rho,\rho\rangle
-\sqrt{\frac{3}{35}}|\omega_1,\omega_1\rangle
-\sqrt{\frac{8}{35}}|\pi,\pi\rangle
\nonumber \\
(+)~~&
|\mathbf{189};\mathbf{1_1}\rangle
&=
\phantom{+}\sqrt{\frac{1}{60}}|\rho,\rho\rangle
-\sqrt{\frac{8}{15}}|\omega_1,\omega_1\rangle
+\sqrt{\frac{9}{20}}|\pi,\pi\rangle
\nonumber \\
(+)~~&
|\mathbf{405};\mathbf{1_1}\rangle
&=
\phantom{+}\sqrt{\frac{25}{84}}|\rho,\rho\rangle
+\sqrt{\frac{8}{21}}|\omega_1,\omega_1\rangle
+\sqrt{\frac{9}{28}}|\pi,\pi\rangle
\end{eqnarray}
\begin{eqnarray}
(+)~~&
|\mathbf{35_s};\mathbf{1_3}\rangle
&=
\phantom{+}|\rho,\pi\rangle_S
\nonumber \\
(-)~~&
|\mathbf{35_a};\mathbf{1_3}\rangle
&=
\phantom{+}\sqrt{\frac{8}{9}}|\rho,\rho\rangle
-\sqrt{\frac{1}{9}}|\omega_1,\omega_1\rangle
\nonumber \\
(-)~~&
|\mathbf{280};\mathbf{1_3}\rangle
&=
\phantom{+}\sqrt{\frac{1}{18}}|\rho,\rho\rangle
+\sqrt{\frac{1}{2}}|\rho,\pi\rangle_A
+\sqrt{\frac{4}{9}}|\omega_1,\omega_1\rangle
\nonumber \\
(-)~~&
|\mathbf{280^*};\mathbf{1_3}\rangle
&=
\phantom{+}\sqrt{\frac{1}{18}}|\rho,\rho\rangle
-\sqrt{\frac{1}{2}}|\rho,\pi\rangle_A
+\sqrt{\frac{4}{9}}|\omega_1,\omega_1\rangle
\end{eqnarray}
\begin{eqnarray}
(+)~~&
|\mathbf{189};\mathbf{1_5}\rangle
&=
\phantom{+}\sqrt{\frac{2}{3}}|\rho,\rho\rangle
+\sqrt{\frac{1}{3}}|\omega_1,\omega_1\rangle
\nonumber \\
(+)~~&
|\mathbf{405};\mathbf{1_5}\rangle
&=
\phantom{+}\sqrt{\frac{1}{3}}|\rho,\rho\rangle
-\sqrt{\frac{2}{3}}|\omega_1,\omega_1\rangle
\end{eqnarray}
\begin{eqnarray}
(+)~~&
|\mathbf{35_s};\mathbf{8_1}\rangle
&=
\phantom{+}\sqrt{\frac{15}{32}}|(\rho,\rho)_s\rangle
+\sqrt{\frac{3}{8}}|\rho,\omega_1\rangle_S
-\sqrt{\frac{5}{32}}|(\pi,\pi)_s\rangle
\nonumber \\
(-)~~&
|\mathbf{35_a};\mathbf{8_1}\rangle
&=
\phantom{+}\sqrt{\frac{3}{4}}|(\rho,\rho)_a\rangle
-\sqrt{\frac{1}{4}}|(\pi,\pi)_a\rangle
\nonumber \\
(+)~~&
|\mathbf{189};\mathbf{8_1}\rangle
&=
\phantom{+}\sqrt{\frac{1}{48}}|(\rho,\rho)_s\rangle
-\sqrt{\frac{5}{12}}|\rho,\omega_1\rangle_S
-\sqrt{\frac{9}{16}}|(\pi,\pi)_s\rangle
\nonumber \\
(-)~~&
|\mathbf{280};\mathbf{8_1}\rangle
&=
\phantom{+}\sqrt{\frac{1}{8}}|(\rho,\rho)_a\rangle
+\sqrt{\frac{1}{2}}|\rho,\omega_1\rangle_A
+\sqrt{\frac{3}{8}}|(\pi,\pi)_a\rangle
\nonumber \\
(-)~~&
|\mathbf{280^*};\mathbf{8_1}\rangle
&=
\phantom{+}\sqrt{\frac{1}{8}}|(\rho,\rho)_a\rangle
-\sqrt{\frac{1}{2}}|\rho,\omega_1\rangle_A
+\sqrt{\frac{3}{8}}|(\pi,\pi)_a\rangle
\nonumber \\
(+)~~&
|\mathbf{405};\mathbf{8_1}\rangle
&=
\phantom{+}\sqrt{\frac{49}{96}}|(\rho,\rho)_s\rangle
-\sqrt{\frac{5}{24}}|\rho,\omega_1\rangle_S
+\sqrt{\frac{9}{32}}|(\pi,\pi)_s\rangle
\end{eqnarray}
\begin{eqnarray}
(+)~~&
|\mathbf{35_s};\mathbf{8_3}\rangle
&=
\phantom{+}\sqrt{\frac{9}{16}}|(\rho,\rho)_a\rangle
+\sqrt{\frac{5}{16}}|(\rho,\pi)_s\rangle_S
+\sqrt{\frac{1}{8}}|\omega_1,\pi\rangle_S
\nonumber \\
(-)~~&
|\mathbf{35_a};\mathbf{8_3}\rangle
&=
\phantom{+}\sqrt{\frac{5}{18}}|(\rho,\rho)_s\rangle
+\sqrt{\frac{2}{9}}|\rho,\omega_1\rangle_S
+\sqrt{\frac{1}{2}}|(\rho,\pi)_a\rangle_S
\nonumber \\
(+)~~&
|\mathbf{189};\mathbf{8_{s,3}}\rangle
&=
\phantom{+}\sqrt{\frac{1}{8}}|(\rho,\rho)_a\rangle
-\sqrt{\frac{5}{8}}|(\rho,\pi)_s\rangle_S
+\sqrt{\frac{1}{4}}|\omega_1,\pi\rangle_S
\nonumber \\
(+)~~&
|\mathbf{189};\mathbf{8_{a,3}}\rangle
&=
\phantom{+}\sqrt{\frac{1}{2}}|\rho,\omega_1\rangle_A
-\sqrt{\frac{1}{2}}|(\rho,\pi)_a\rangle_A
\nonumber \\
(-)~~&
|\mathbf{280};\mathbf{8_{s,3}}\rangle
&=
\phantom{+}\sqrt{\frac{13}{36}}|(\rho,\rho)_s\rangle
-\sqrt{\frac{5}{117}}|\rho,\omega_1\rangle_S
+\sqrt{\frac{4}{13}}|(\rho,\pi)_s\rangle_A
-\sqrt{\frac{5}{52}}|(\rho,\pi)_a\rangle_S
\nonumber \\ &&
-\sqrt{\frac{5}{26}}|\omega_1,\pi\rangle_A
\nonumber \\
(-)~~&
|\mathbf{280};\mathbf{8_{a,3}}\rangle
&=
\phantom{+}\sqrt{\frac{9}{26}}|\rho,\omega_1\rangle_S
-\sqrt{\frac{5}{26}}|(\rho,\pi)_s\rangle_A
-\sqrt{\frac{2}{13}}|(\rho,\pi)_a\rangle_S
-\sqrt{\frac{4}{13}}|\omega_1,\pi\rangle_A
\nonumber \\
(-)~~&
|\mathbf{280^*};\mathbf{8_{s,3}}\rangle
&=
\phantom{+}\sqrt{\frac{13}{36}}|(\rho,\rho)_s\rangle
-\sqrt{\frac{5}{117}}|\rho,\omega_1\rangle_S
-\sqrt{\frac{4}{13}}|(\rho,\pi)_s\rangle_A
-\sqrt{\frac{5}{52}}|(\rho,\pi)_a\rangle_S
\nonumber \\ &&
+\sqrt{\frac{5}{26}}|\omega_1,\pi\rangle_A
\nonumber \\
(-)~~&
|\mathbf{280^*};\mathbf{8_{a,3}}\rangle
&=
\phantom{+}\sqrt{\frac{9}{26}}|\rho,\omega_1\rangle_S
+\sqrt{\frac{5}{26}}|(\rho,\pi)_s\rangle_A
-\sqrt{\frac{2}{13}}|(\rho,\pi)_a\rangle_S
+\sqrt{\frac{4}{13}}|\omega_1,\pi\rangle_A
\nonumber \\
(+)~~&
|\mathbf{405};\mathbf{8_{s,3}}\rangle
&=
\phantom{+}\sqrt{\frac{5}{16}}|(\rho,\rho)_a\rangle
-\sqrt{\frac{1}{16}}|(\rho,\pi)_s\rangle_S
-\sqrt{\frac{5}{8}}|\omega_1,\pi\rangle_S
\nonumber \\
(+)~~&
|\mathbf{405};\mathbf{8_{a,3}}\rangle
&=
\phantom{+}\sqrt{\frac{1}{2}}|\rho,\omega_1\rangle_A
+\sqrt{\frac{1}{2}}|(\rho,\pi)_a\rangle_A
\end{eqnarray}
\begin{eqnarray}
(+)~~&
|\mathbf{189};\mathbf{8_5}\rangle
&=
\phantom{+}\sqrt{\frac{5}{6}}|(\rho,\rho)_s\rangle
-\sqrt{\frac{1}{6}}|\rho,\omega_1\rangle_S
\nonumber \\
(-)~~&
|\mathbf{280};\mathbf{8_5}\rangle
&=
\phantom{+}\sqrt{\frac{1}{2}}|(\rho,\rho)_a\rangle
-\sqrt{\frac{1}{2}}|\rho,\omega_1\rangle_A
\nonumber \\
(-)~~&
|\mathbf{280^*};\mathbf{8_5}\rangle
&=
\phantom{+}\sqrt{\frac{1}{2}}|(\rho,\rho)_a\rangle
+\sqrt{\frac{1}{2}}|\rho,\omega_1\rangle_A
\nonumber \\
(+)~~&
|\mathbf{405};\mathbf{8_5}\rangle
&=
\phantom{+}\sqrt{\frac{1}{6}}|(\rho,\rho)_s\rangle
+\sqrt{\frac{5}{6}}|\rho,\omega_1\rangle_S
\end{eqnarray}
\begin{eqnarray}
(-)~~&
|\mathbf{280};\mathbf{10_1}\rangle
&=
\phantom{+}\sqrt{\frac{1}{4}}|\rho,\rho\rangle
-\sqrt{\frac{3}{4}}|\pi,\pi\rangle
\nonumber \\
(-)~~&
|\mathbf{280^*};\mathbf{10_1}\rangle
&=
\phantom{+}\sqrt{\frac{3}{4}}|\rho,\rho\rangle
+\sqrt{\frac{1}{4}}|\pi,\pi\rangle
\end{eqnarray}
\begin{eqnarray}
(+)~~&
|\mathbf{189};\mathbf{10_3}\rangle
&=
\phantom{+}\sqrt{\frac{1}{2}}|\rho,\rho\rangle
+\sqrt{\frac{1}{2}}|\rho,\pi\rangle_A
\nonumber \\
(-)~~&
|\mathbf{280};\mathbf{10_3}\rangle
&=
\phantom{+}|\rho,\pi\rangle_S
\nonumber \\
(+)~~&
|\mathbf{405};\mathbf{10_3}\rangle
&=
\phantom{+}\sqrt{\frac{1}{2}}|\rho,\rho\rangle
-\sqrt{\frac{1}{2}}|\rho,\pi\rangle_A
\end{eqnarray}
\begin{eqnarray}
(-)~~&
|\mathbf{280};\mathbf{10_5}\rangle
&=
\phantom{+}|\rho,\rho\rangle
\end{eqnarray}
\begin{eqnarray}
(-)~~&
|\mathbf{280};\mathbf{10^*_1}\rangle
&=
\phantom{+}\sqrt{\frac{3}{4}}|\rho,\rho\rangle
+\sqrt{\frac{1}{4}}|\pi,\pi\rangle
\nonumber \\
(-)~~&
|\mathbf{280^*};\mathbf{10^*_1}\rangle
&=
\phantom{+}\sqrt{\frac{1}{4}}|\rho,\rho\rangle
-\sqrt{\frac{3}{4}}|\pi,\pi\rangle
\end{eqnarray}
\begin{eqnarray}
(+)~~&
|\mathbf{189};\mathbf{10^*_3}\rangle
&=
\phantom{+}\sqrt{\frac{1}{2}}|\rho,\rho\rangle
-\sqrt{\frac{1}{2}}|\rho,\pi\rangle_A
\nonumber \\
(-)~~&
|\mathbf{280^*};\mathbf{10^*_3}\rangle
&=
\phantom{+}|\rho,\pi\rangle_S
\nonumber \\
(+)~~&
|\mathbf{405};\mathbf{10^*_3}\rangle
&=
\phantom{+}\sqrt{\frac{1}{2}}|\rho,\rho\rangle
+\sqrt{\frac{1}{2}}|\rho,\pi\rangle_A
\end{eqnarray}
\begin{eqnarray}
(-)~~&
|\mathbf{280^*};\mathbf{10^*_5}\rangle
&=
\phantom{+}|\rho,\rho\rangle
\end{eqnarray}
\begin{eqnarray}
(+)~~&
|\mathbf{189};\mathbf{27_1}\rangle
&=
\phantom{+}\sqrt{\frac{3}{4}}|\rho,\rho\rangle
+\sqrt{\frac{1}{4}}|\pi,\pi\rangle
\nonumber \\
(+)~~&
|\mathbf{405};\mathbf{27_1}\rangle
&=
\phantom{+}\sqrt{\frac{1}{4}}|\rho,\rho\rangle
-\sqrt{\frac{3}{4}}|\pi,\pi\rangle
\end{eqnarray}
\begin{eqnarray}
(-)~~&
|\mathbf{280};\mathbf{27_3}\rangle
&=
\phantom{+}\sqrt{\frac{1}{2}}|\rho,\rho\rangle
-\sqrt{\frac{1}{2}}|\rho,\pi\rangle_A
\nonumber \\
(-)~~&
|\mathbf{280^*};\mathbf{27_3}\rangle
&=
\phantom{+}\sqrt{\frac{1}{2}}|\rho,\rho\rangle
+\sqrt{\frac{1}{2}}|\rho,\pi\rangle_A
\nonumber \\
(+)~~&
|\mathbf{405};\mathbf{27_3}\rangle
&=
\phantom{+}|\rho,\pi\rangle_S
\end{eqnarray}
\begin{eqnarray}
(+)~~&
|\mathbf{405};\mathbf{27_5}\rangle
&=
\phantom{+}|\rho,\rho\rangle
\end{eqnarray}
\subsection{SU(6): $ \mathbf{56} \otimes \mathbf{35}$} \par
\begin{eqnarray}
(-)~~&
|\mathbf{70};\mathbf{1_2}\rangle
&=
\phantom{+}\sqrt{\frac{3}{4}}|\Sigma,\rho\rangle
-\sqrt{\frac{1}{4}}|\Sigma,\pi\rangle
\nonumber \\
(-)~~&
|\mathbf{1134};\mathbf{1_2}\rangle
&=
\phantom{+}\sqrt{\frac{1}{4}}|\Sigma,\rho\rangle
+\sqrt{\frac{3}{4}}|\Sigma,\pi\rangle
\end{eqnarray}
\begin{eqnarray}
(+)~~&
|\mathbf{1134};\mathbf{1_4}\rangle
&=
\phantom{+}|\Sigma,\rho\rangle
\end{eqnarray}
\begin{eqnarray}
(+)~~&
|\mathbf{56};\mathbf{8_2}\rangle
&=
\phantom{+}\sqrt{\frac{4}{9}}|\Delta,\rho\rangle
+\sqrt{\frac{2}{9}}|(\Sigma,\rho)_s\rangle
-\sqrt{\frac{8}{45}}|(\Sigma,\rho)_a\rangle
+\sqrt{\frac{1}{45}}|\Sigma,\omega_1\rangle
\nonumber \\ &&
+\sqrt{\frac{2}{15}}|(\Sigma,\pi)_a\rangle
\nonumber \\
(+)~~&
|\mathbf{70};\mathbf{8_2}\rangle
&=
\phantom{+}\sqrt{\frac{5}{12}}|\Delta,\rho\rangle
-\sqrt{\frac{5}{96}}|(\Sigma,\rho)_s\rangle
+\sqrt{\frac{25}{96}}|(\Sigma,\rho)_a\rangle
-\sqrt{\frac{1}{12}}|\Sigma,\omega_1\rangle
\nonumber \\ &&
+\sqrt{\frac{5}{32}}|(\Sigma,\pi)_s\rangle
-\sqrt{\frac{1}{32}}|(\Sigma,\pi)_a\rangle
\nonumber \\
(+)~~&
|\mathbf{700};\mathbf{8_2}\rangle
&=
\phantom{+}\sqrt{\frac{1}{18}}|\Delta,\rho\rangle
-\sqrt{\frac{49}{144}}|(\Sigma,\rho)_s\rangle
+\sqrt{\frac{5}{144}}|(\Sigma,\rho)_a\rangle
+\sqrt{\frac{5}{18}}|\Sigma,\omega_1\rangle
\nonumber \\ &&
-\sqrt{\frac{3}{16}}|(\Sigma,\pi)_s\rangle
+\sqrt{\frac{5}{48}}|(\Sigma,\pi)_a\rangle
\nonumber \\
(+)~~&
|\mathbf{1134};\mathbf{8_{s,2}}\rangle
&=
\phantom{+}\sqrt{\frac{1}{12}}|\Delta,\rho\rangle
-\sqrt{\frac{1}{96}}|(\Sigma,\rho)_s\rangle
-\sqrt{\frac{49}{480}}|(\Sigma,\rho)_a\rangle
-\sqrt{\frac{1}{60}}|\Sigma,\omega_1\rangle
\nonumber \\ &&
-\sqrt{\frac{9}{32}}|(\Sigma,\pi)_s\rangle
-\sqrt{\frac{81}{160}}|(\Sigma,\pi)_a\rangle
\nonumber \\
(-)~~&
|\mathbf{1134};\mathbf{8_{a,2}}\rangle
&=
\phantom{+}\sqrt{\frac{3}{8}}|(\Sigma,\rho)_s\rangle
+\sqrt{\frac{49}{120}}|(\Sigma,\rho)_a\rangle
+\sqrt{\frac{1}{15}}|\Sigma,\omega_1\rangle
-\sqrt{\frac{1}{8}}|(\Sigma,\pi)_s\rangle
\nonumber \\ &&
-\sqrt{\frac{1}{40}}|(\Sigma,\pi)_a\rangle
\nonumber \\
(+)~~&
|\mathbf{1134};\mathbf{8_{b,2}}\rangle
&=
\phantom{+}\sqrt{\frac{1}{60}}|(\Sigma,\rho)_a\rangle
-\sqrt{\frac{8}{15}}|\Sigma,\omega_1\rangle
-\sqrt{\frac{1}{4}}|(\Sigma,\pi)_s\rangle
+\sqrt{\frac{1}{5}}|(\Sigma,\pi)_a\rangle
\end{eqnarray}
\begin{eqnarray}
(-)~~&
|\mathbf{70};\mathbf{8_4}\rangle
&=
\phantom{+}\sqrt{\frac{25}{48}}|\Delta,\rho\rangle
-\sqrt{\frac{5}{16}}|\Delta,\pi\rangle
+\sqrt{\frac{5}{48}}|(\Sigma,\rho)_s\rangle
+\sqrt{\frac{1}{48}}|(\Sigma,\rho)_a\rangle
\nonumber \\ &&
-\sqrt{\frac{1}{24}}|\Sigma,\omega_1\rangle
\nonumber \\
(-)~~&
|\mathbf{700};\mathbf{8_4}\rangle
&=
\phantom{+}\sqrt{\frac{5}{24}}|\Delta,\rho\rangle
+\sqrt{\frac{1}{8}}|\Delta,\pi\rangle
+\sqrt{\frac{1}{24}}|(\Sigma,\rho)_s\rangle
-\sqrt{\frac{5}{24}}|(\Sigma,\rho)_a\rangle
\nonumber \\ &&
+\sqrt{\frac{5}{12}}|\Sigma,\omega_1\rangle
\nonumber \\
(-)~~&
|\mathbf{1134};\mathbf{8_{s,4}}\rangle
&=
\phantom{+}\sqrt{\frac{13}{48}}|\Delta,\rho\rangle
+\sqrt{\frac{45}{208}}|\Delta,\pi\rangle
-\sqrt{\frac{245}{624}}|(\Sigma,\rho)_s\rangle
+\sqrt{\frac{25}{624}}|(\Sigma,\rho)_a\rangle
\nonumber \\ &&
-\sqrt{\frac{25}{312}}|\Sigma,\omega_1\rangle
\nonumber \\
(+)~~&
|\mathbf{1134};\mathbf{8_{a,4}}\rangle
&=
\phantom{+}\sqrt{\frac{9}{26}}|\Delta,\pi\rangle
+\sqrt{\frac{6}{13}}|(\Sigma,\rho)_s\rangle
+\sqrt{\frac{5}{78}}|(\Sigma,\rho)_a\rangle
-\sqrt{\frac{5}{39}}|\Sigma,\omega_1\rangle
\nonumber \\
(-)~~&
|\mathbf{1134};\mathbf{8_{b,4}}\rangle
&=
\phantom{+}\sqrt{\frac{2}{3}}|(\Sigma,\rho)_a\rangle
+\sqrt{\frac{1}{3}}|\Sigma,\omega_1\rangle
\end{eqnarray}
\begin{eqnarray}
(+)~~&
|\mathbf{1134};\mathbf{8_6}\rangle
&=
\phantom{+}|\Delta,\rho\rangle
\end{eqnarray}
\begin{eqnarray}
(-)~~&
|\mathbf{70};\mathbf{10_2}\rangle
&=
\phantom{+}\sqrt{\frac{2}{3}}|\Delta,\rho\rangle
-\sqrt{\frac{1}{6}}|\Delta,\omega_1\rangle
+\sqrt{\frac{1}{24}}|\Sigma,\rho\rangle
+\sqrt{\frac{1}{8}}|\Sigma,\pi\rangle
\nonumber \\
(-)~~&
|\mathbf{700};\mathbf{10_2}\rangle
&=
\phantom{+}\sqrt{\frac{1}{6}}|\Delta,\rho\rangle
+\sqrt{\frac{1}{6}}|\Delta,\omega_1\rangle
+\sqrt{\frac{1}{6}}|\Sigma,\rho\rangle
-\sqrt{\frac{1}{2}}|\Sigma,\pi\rangle
\nonumber \\
(-)~~&
|\mathbf{1134};\mathbf{10_{s,2}}\rangle
&=
\phantom{+}\sqrt{\frac{1}{6}}|\Delta,\rho\rangle
+\sqrt{\frac{1}{6}}|\Delta,\omega_1\rangle
-\sqrt{\frac{2}{3}}|\Sigma,\rho\rangle
\nonumber \\
(+)~~&
|\mathbf{1134};\mathbf{10_{a,2}}\rangle
&=
\phantom{+}\sqrt{\frac{1}{2}}|\Delta,\omega_1\rangle
+\sqrt{\frac{1}{8}}|\Sigma,\rho\rangle
+\sqrt{\frac{3}{8}}|\Sigma,\pi\rangle
\end{eqnarray}
\begin{eqnarray}
(+)~~&
|\mathbf{56};\mathbf{10_4}\rangle
&=
\phantom{+}\sqrt{\frac{4}{9}}|\Delta,\rho\rangle
-\sqrt{\frac{1}{9}}|\Delta,\omega_1\rangle
-\sqrt{\frac{4}{15}}|\Delta,\pi\rangle
+\sqrt{\frac{8}{45}}|\Sigma,\rho\rangle
\nonumber \\
(+)~~&
|\mathbf{700};\mathbf{10_4}\rangle
&=
\phantom{+}\sqrt{\frac{1}{72}}|\Delta,\rho\rangle
+\sqrt{\frac{2}{9}}|\Delta,\omega_1\rangle
+\sqrt{\frac{5}{24}}|\Delta,\pi\rangle
+\sqrt{\frac{5}{9}}|\Sigma,\rho\rangle
\nonumber \\
(+)~~&
|\mathbf{1134};\mathbf{10_{s,4}}\rangle
&=
\phantom{+}\sqrt{\frac{13}{24}}|\Delta,\rho\rangle
+\sqrt{\frac{2}{39}}|\Delta,\omega_1\rangle
+\sqrt{\frac{81}{520}}|\Delta,\pi\rangle
-\sqrt{\frac{49}{195}}|\Sigma,\rho\rangle
\nonumber \\
(-)~~&
|\mathbf{1134};\mathbf{10_{a,4}}\rangle
&=
\phantom{+}\sqrt{\frac{8}{13}}|\Delta,\omega_1\rangle
-\sqrt{\frac{24}{65}}|\Delta,\pi\rangle
-\sqrt{\frac{1}{65}}|\Sigma,\rho\rangle
\end{eqnarray}
\begin{eqnarray}
(-)~~&
|\mathbf{700};\mathbf{10_6}\rangle
&=
\phantom{+}\sqrt{\frac{1}{2}}|\Delta,\rho\rangle
-\sqrt{\frac{1}{2}}|\Delta,\omega_1\rangle
\nonumber \\
(-)~~&
|\mathbf{1134};\mathbf{10_6}\rangle
&=
\phantom{+}\sqrt{\frac{1}{2}}|\Delta,\rho\rangle
+\sqrt{\frac{1}{2}}|\Delta,\omega_1\rangle
\end{eqnarray}
\begin{eqnarray}
(+)~~&
|\mathbf{700};\mathbf{10^*_2}\rangle
&=
\phantom{+}\sqrt{\frac{3}{4}}|\Sigma,\rho\rangle
+\sqrt{\frac{1}{4}}|\Sigma,\pi\rangle
\nonumber \\
(+)~~&
|\mathbf{1134};\mathbf{10^*_2}\rangle
&=
\phantom{+}\sqrt{\frac{1}{4}}|\Sigma,\rho\rangle
-\sqrt{\frac{3}{4}}|\Sigma,\pi\rangle
\end{eqnarray}
\begin{eqnarray}
(-)~~&
|\mathbf{1134};\mathbf{10^*_4}\rangle
&=
\phantom{+}|\Sigma,\rho\rangle
\end{eqnarray}
\begin{eqnarray}
(-)~~&
|\mathbf{700};\mathbf{27_2}\rangle
&=
\phantom{+}\sqrt{\frac{1}{3}}|\Delta,\rho\rangle
-\sqrt{\frac{1}{6}}|\Sigma,\rho\rangle
+\sqrt{\frac{1}{2}}|\Sigma,\pi\rangle
\nonumber \\
(-)~~&
|\mathbf{1134};\mathbf{27_{s,2}}\rangle
&=
\phantom{+}\sqrt{\frac{2}{3}}|\Delta,\rho\rangle
+\sqrt{\frac{1}{12}}|\Sigma,\rho\rangle
-\sqrt{\frac{1}{4}}|\Sigma,\pi\rangle
\nonumber \\
(-)~~&
|\mathbf{1134};\mathbf{27_{a,2}}\rangle
&=
\phantom{+}\sqrt{\frac{3}{4}}|\Sigma,\rho\rangle
+\sqrt{\frac{1}{4}}|\Sigma,\pi\rangle
\end{eqnarray}
\begin{eqnarray}
(+)~~&
|\mathbf{700};\mathbf{27_4}\rangle
&=
\phantom{+}\sqrt{\frac{5}{24}}|\Delta,\rho\rangle
+\sqrt{\frac{1}{8}}|\Delta,\pi\rangle
-\sqrt{\frac{2}{3}}|\Sigma,\rho\rangle
\nonumber \\
(+)~~&
|\mathbf{1134};\mathbf{27_{s,4}}\rangle
&=
\phantom{+}\sqrt{\frac{19}{24}}|\Delta,\rho\rangle
-\sqrt{\frac{5}{152}}|\Delta,\pi\rangle
+\sqrt{\frac{10}{57}}|\Sigma,\rho\rangle
\nonumber \\
(-)~~&
|\mathbf{1134};\mathbf{27_{a,4}}\rangle
&=
\phantom{+}\sqrt{\frac{16}{19}}|\Delta,\pi\rangle
+\sqrt{\frac{3}{19}}|\Sigma,\rho\rangle
\end{eqnarray}
\begin{eqnarray}
(-)~~&
|\mathbf{1134};\mathbf{27_6}\rangle
&=
\phantom{+}|\Delta,\rho\rangle
\end{eqnarray}
\begin{eqnarray}
(+)~~&
|\mathbf{1134};\mathbf{35_2}\rangle
&=
\phantom{+}|\Delta,\rho\rangle
\end{eqnarray}
\begin{eqnarray}
(-)~~&
|\mathbf{700};\mathbf{35_4}\rangle
&=
\phantom{+}\sqrt{\frac{3}{8}}|\Delta,\rho\rangle
-\sqrt{\frac{5}{8}}|\Delta,\pi\rangle
\nonumber \\
(-)~~&
|\mathbf{1134};\mathbf{35_4}\rangle
&=
\phantom{+}\sqrt{\frac{5}{8}}|\Delta,\rho\rangle
+\sqrt{\frac{3}{8}}|\Delta,\pi\rangle
\end{eqnarray}
\begin{eqnarray}
(+)~~&
|\mathbf{700};\mathbf{35_6}\rangle
&=
\phantom{+}|\Delta,\rho\rangle
\end{eqnarray}
\section{Tables of scalar factors of SU(4)}
\begin{eqnarray}
|\pi\rangle &=& |\mathbf{15};\mathbf{8},0\rangle
,
\quad
|D\rangle = |\mathbf{15};\mathbf{3^*},1\rangle
,
\quad
|\bar{D}\rangle = |\mathbf{15};\mathbf{3},-1\rangle
,
\quad
|\eta_c\rangle = |\mathbf{15};\mathbf{1},0\rangle
,
\nonumber\\
|\Sigma\rangle &=& |\mathbf{20};\mathbf{8},0\rangle
,
\quad
|\Sigma_c\rangle = |\mathbf{20};\mathbf{6},1\rangle
,
\quad
|\Xi_c\rangle = |\mathbf{20};\mathbf{3^*},1\rangle
,
\quad
|\Xi_{cc}\rangle = |\mathbf{20};\mathbf{3},2\rangle
,
\nonumber\\
|\Delta\rangle &=& |\mathbf{20^\prime};\mathbf{10},0\rangle
,
\quad
|\Sigma_c^*\rangle = |\mathbf{20^\prime};\mathbf{6},1\rangle
,
\quad
|\Xi^*_{cc}\rangle = |\mathbf{20^\prime};\mathbf{3},2\rangle
,
\quad
|\Omega_{ccc}\rangle = |\mathbf{20^\prime};\mathbf{1},3\rangle
.
\nonumber
\end{eqnarray}

\subsection{SU(4): $ \mathbf{15} \otimes \mathbf{15}$} \par
\begin{eqnarray}
(+)~~&
|\mathbf{1};\mathbf{1},0\rangle
&=
\phantom{+}\sqrt{\frac{8}{15}}|\pi,\pi\rangle
-\sqrt{\frac{2}{5}}|D,{\bar D}\rangle_A
-\sqrt{\frac{1}{15}}|\eta_c,\eta_c\rangle
\nonumber \\
(+)~~&
|\mathbf{15_s};\mathbf{1},0\rangle
&=
\phantom{+}\sqrt{\frac{4}{9}}|\pi,\pi\rangle
+\sqrt{\frac{1}{3}}|D,{\bar D}\rangle_A
+\sqrt{\frac{2}{9}}|\eta_c,\eta_c\rangle
\nonumber \\
(-)~~&
|\mathbf{15_a};\mathbf{1},0\rangle
&=
\phantom{+}|D,{\bar D}\rangle_S
\nonumber \\
(+)~~&
|\mathbf{84};\mathbf{1},0\rangle
&=
-\sqrt{\frac{1}{45}}|\pi,\pi\rangle
-\sqrt{\frac{4}{15}}|D,{\bar D}\rangle_A
+\sqrt{\frac{32}{45}}|\eta_c,\eta_c\rangle
\end{eqnarray}
\begin{eqnarray}
(+)~~&
|\mathbf{15_s};\mathbf{3},-1\rangle
&=
\phantom{+}\sqrt{\frac{8}{9}}|\pi,{\bar D}\rangle_A
+\sqrt{\frac{1}{9}}|{\bar D},\eta_c\rangle_S
\nonumber \\
(-)~~&
|\mathbf{15_a};\mathbf{3},-1\rangle
&=
\phantom{+}\sqrt{\frac{2}{3}}|\pi,{\bar D}\rangle_S
+\sqrt{\frac{1}{3}}|{\bar D},\eta_c\rangle_A
\nonumber \\
(-)~~&
|\mathbf{45^*};\mathbf{3},-1\rangle
&=
-\sqrt{\frac{1}{3}}|\pi,{\bar D}\rangle_S
+\sqrt{\frac{2}{3}}|{\bar D},\eta_c\rangle_A
\nonumber \\
(+)~~&
|\mathbf{84};\mathbf{3},-1\rangle
&=
-\sqrt{\frac{1}{9}}|\pi,{\bar D}\rangle_A
+\sqrt{\frac{8}{9}}|{\bar D},\eta_c\rangle_S
\end{eqnarray}
\begin{eqnarray}
(-)~~&
|\mathbf{45};\mathbf{3},2\rangle
&=
\phantom{+}|D,D\rangle
\end{eqnarray}
\begin{eqnarray}
(-)~~&
|\mathbf{45^*};\mathbf{3^*},-2\rangle
&=
\phantom{+}|{\bar D},{\bar D}\rangle
\end{eqnarray}
\begin{eqnarray}
(+)~~&
|\mathbf{15_s};\mathbf{3^*},1\rangle
&=
-\sqrt{\frac{8}{9}}|\pi,D\rangle_A
+\sqrt{\frac{1}{9}}|D,\eta_c\rangle_S
\nonumber \\
(-)~~&
|\mathbf{15_a};\mathbf{3^*},1\rangle
&=
\phantom{+}\sqrt{\frac{2}{3}}|\pi,D\rangle_S
-\sqrt{\frac{1}{3}}|D,\eta_c\rangle_A
\nonumber \\
(-)~~&
|\mathbf{45};\mathbf{3^*},1\rangle
&=
\phantom{+}\sqrt{\frac{1}{3}}|\pi,D\rangle_S
+\sqrt{\frac{2}{3}}|D,\eta_c\rangle_A
\nonumber \\
(+)~~&
|\mathbf{84};\mathbf{3^*},1\rangle
&=
\phantom{+}\sqrt{\frac{1}{9}}|\pi,D\rangle_A
+\sqrt{\frac{8}{9}}|D,\eta_c\rangle_S
\end{eqnarray}
\begin{eqnarray}
(+)~~&
|\mathbf{84};\mathbf{6},-2\rangle
&=
\phantom{+}|{\bar D},{\bar D}\rangle
\end{eqnarray}
\begin{eqnarray}
(+)~~&
|\mathbf{20^{\prime\prime}};\mathbf{6},1\rangle
&=
\phantom{+}|\pi,D\rangle_A
\nonumber \\
(-)~~&
|\mathbf{45};\mathbf{6},1\rangle
&=
\phantom{+}|\pi,D\rangle_S
\end{eqnarray}
\begin{eqnarray}
(+)~~&
|\mathbf{20^{\prime\prime}};\mathbf{6^*},-1\rangle
&=
\phantom{+}|\pi,{\bar D}\rangle_A
\nonumber \\
(-)~~&
|\mathbf{45^*};\mathbf{6^*},-1\rangle
&=
\phantom{+}|\pi,{\bar D}\rangle_S
\end{eqnarray}
\begin{eqnarray}
(+)~~&
|\mathbf{84};\mathbf{6^*},2\rangle
&=
\phantom{+}|D,D\rangle
\end{eqnarray}
\begin{eqnarray}
(+)~~&
|\mathbf{15_s};\mathbf{8},0\rangle
&=
\phantom{+}\sqrt{\frac{5}{9}}|(\pi,\pi)_s\rangle
-\sqrt{\frac{1}{9}}|\pi,\eta_c\rangle_S
+\sqrt{\frac{1}{3}}|D,{\bar D}\rangle_S
\nonumber \\
(-)~~&
|\mathbf{15_a};\mathbf{8},0\rangle
&=
\phantom{+}\sqrt{\frac{3}{4}}|(\pi,\pi)_a\rangle
-\sqrt{\frac{1}{4}}|D,{\bar D}\rangle_A
\nonumber \\
(+)~~&
|\mathbf{20^{\prime\prime}};\mathbf{8},0\rangle
&=
\phantom{+}\sqrt{\frac{5}{12}}|(\pi,\pi)_s\rangle
+\sqrt{\frac{1}{3}}|\pi,\eta_c\rangle_S
-\sqrt{\frac{1}{4}}|D,{\bar D}\rangle_S
\nonumber \\
(-)~~&
|\mathbf{45};\mathbf{8},0\rangle
&=
\phantom{+}\sqrt{\frac{1}{8}}|(\pi,\pi)_a\rangle
+\sqrt{\frac{1}{2}}|\pi,\eta_c\rangle_A
+\sqrt{\frac{3}{8}}|D,{\bar D}\rangle_A
\nonumber \\
(-)~~&
|\mathbf{45^*};\mathbf{8},0\rangle
&=
-\sqrt{\frac{1}{8}}|(\pi,\pi)_a\rangle
+\sqrt{\frac{1}{2}}|\pi,\eta_c\rangle_A
-\sqrt{\frac{3}{8}}|D,{\bar D}\rangle_A
\nonumber \\
(+)~~&
|\mathbf{84};\mathbf{8},0\rangle
&=
-\sqrt{\frac{1}{36}}|(\pi,\pi)_s\rangle
+\sqrt{\frac{5}{9}}|\pi,\eta_c\rangle_S
+\sqrt{\frac{5}{12}}|D,{\bar D}\rangle_S
\end{eqnarray}
\begin{eqnarray}
(-)~~&
|\mathbf{45};\mathbf{10},0\rangle
&=
\phantom{+}|\pi,\pi\rangle
\end{eqnarray}
\begin{eqnarray}
(-)~~&
|\mathbf{45^*};\mathbf{10^*},0\rangle
&=
\phantom{+}|\pi,\pi\rangle
\end{eqnarray}
\begin{eqnarray}
(-)~~&
|\mathbf{45};\mathbf{15},-1\rangle
&=
\phantom{+}|\pi,{\bar D}\rangle_A
\nonumber \\
(+)~~&
|\mathbf{84};\mathbf{15},-1\rangle
&=
\phantom{+}|\pi,{\bar D}\rangle_S
\end{eqnarray}
\begin{eqnarray}
(-)~~&
|\mathbf{45^*};\mathbf{15^*},1\rangle
&=
\phantom{+}|\pi,D\rangle_A
\nonumber \\
(+)~~&
|\mathbf{84};\mathbf{15^*},1\rangle
&=
\phantom{+}|\pi,D\rangle_S
\end{eqnarray}
\begin{eqnarray}
(+)~~&
|\mathbf{84};\mathbf{27},0\rangle
&=
\phantom{+}|\pi,\pi\rangle
\end{eqnarray}
\subsection{SU(4): $ \mathbf{20} \otimes \mathbf{15}$} \par
\begin{eqnarray}
(-)~~&
|\mathbf{4^*};\mathbf{1},0\rangle
&=
-\sqrt{\frac{4}{5}}|\Sigma,\pi\rangle
+\sqrt{\frac{1}{5}}|\Xi_c,{\bar D}\rangle
\nonumber \\
(+)~~&
|\mathbf{36^*};\mathbf{1},0\rangle
&=
-\sqrt{\frac{1}{5}}|\Sigma,\pi\rangle
-\sqrt{\frac{4}{5}}|\Xi_c,{\bar D}\rangle
\end{eqnarray}
\begin{eqnarray}
(-)~~&
|\mathbf{20^\prime};\mathbf{1},3\rangle
&=
-|\Xi_{cc},D\rangle
\end{eqnarray}
\begin{eqnarray}
(+)~~&
|\mathbf{36^*};\mathbf{3},-1\rangle
&=
-|\Sigma,{\bar D}\rangle
\end{eqnarray}
\begin{eqnarray}
(+)~~&
|\mathbf{20_s};\mathbf{3},2\rangle
&=
-\sqrt{\frac{121}{234}}|\Sigma_c,D\rangle
+\sqrt{\frac{1}{39}}|\Xi_c,D\rangle
+\sqrt{\frac{289}{702}}|\Xi_{cc},\pi\rangle
+\sqrt{\frac{16}{351}}|\Xi_{cc},\eta_c\rangle
\nonumber \\
(-)~~&
|\mathbf{20_a};\mathbf{3},2\rangle
&=
\phantom{+}\sqrt{\frac{8}{39}}|\Sigma_c,D\rangle
+\sqrt{\frac{4}{13}}|\Xi_c,D\rangle
+\sqrt{\frac{32}{117}}|\Xi_{cc},\pi\rangle
-\sqrt{\frac{25}{117}}|\Xi_{cc},\eta_c\rangle
\nonumber \\
(-)~~&
|\mathbf{20^\prime};\mathbf{3},2\rangle
&=
-\sqrt{\frac{2}{9}}|\Sigma_c,D\rangle
+\sqrt{\frac{1}{3}}|\Xi_c,D\rangle
-\sqrt{\frac{8}{27}}|\Xi_{cc},\pi\rangle
-\sqrt{\frac{4}{27}}|\Xi_{cc},\eta_c\rangle
\nonumber \\
(+)~~&
|\mathbf{140};\mathbf{3},2\rangle
&=
\phantom{+}\sqrt{\frac{1}{18}}|\Sigma_c,D\rangle
+\sqrt{\frac{1}{3}}|\Xi_c,D\rangle
-\sqrt{\frac{1}{54}}|\Xi_{cc},\pi\rangle
+\sqrt{\frac{16}{27}}|\Xi_{cc},\eta_c\rangle
\end{eqnarray}
\begin{eqnarray}
(-)~~&
|\mathbf{4^*};\mathbf{3^*},1\rangle
&=
\phantom{+}\sqrt{\frac{4}{15}}|\Sigma,D\rangle
-\sqrt{\frac{2}{5}}|\Sigma_c,\pi\rangle
-\sqrt{\frac{2}{45}}|\Xi_c,\pi\rangle
-\sqrt{\frac{4}{45}}|\Xi_c,\eta_c\rangle
+\sqrt{\frac{1}{5}}|\Xi_{cc},{\bar D}\rangle
\nonumber \\
(+)~~&
|\mathbf{20_s};\mathbf{3^*},1\rangle
&=
-\sqrt{\frac{9}{52}}|\Sigma,D\rangle
-\sqrt{\frac{13}{24}}|\Sigma_c,\pi\rangle
+\sqrt{\frac{49}{312}}|\Xi_c,\pi\rangle
+\sqrt{\frac{4}{39}}|\Xi_c,\eta_c\rangle
-\sqrt{\frac{1}{39}}|\Xi_{cc},{\bar D}\rangle
\nonumber \\
(-)~~&
|\mathbf{20_a};\mathbf{3^*},1\rangle
&=
\phantom{+}\sqrt{\frac{16}{39}}|\Sigma,D\rangle
+\sqrt{\frac{32}{117}}|\Xi_c,\pi\rangle
-\sqrt{\frac{1}{117}}|\Xi_c,\eta_c\rangle
-\sqrt{\frac{4}{13}}|\Xi_{cc},{\bar D}\rangle
\nonumber \\
(+)~~&
|\mathbf{36^*};\mathbf{3^*},1\rangle
&=
\phantom{+}\sqrt{\frac{1}{40}}|\Sigma,D\rangle
-\sqrt{\frac{3}{80}}|\Sigma_c,\pi\rangle
-\sqrt{\frac{121}{240}}|\Xi_c,\pi\rangle
+\sqrt{\frac{2}{15}}|\Xi_c,\eta_c\rangle
-\sqrt{\frac{3}{10}}|\Xi_{cc},{\bar D}\rangle
\nonumber \\
(+)~~&
|\mathbf{140};\mathbf{3^*},1\rangle
&=
\phantom{+}\sqrt{\frac{1}{8}}|\Sigma,D\rangle
+\sqrt{\frac{1}{48}}|\Sigma_c,\pi\rangle
+\sqrt{\frac{1}{48}}|\Xi_c,\pi\rangle
+\sqrt{\frac{2}{3}}|\Xi_c,\eta_c\rangle
+\sqrt{\frac{1}{6}}|\Xi_{cc},{\bar D}\rangle
\end{eqnarray}
\begin{eqnarray}
(+)~~&
|\mathbf{20_s};\mathbf{6},1\rangle
&=
\phantom{+}\sqrt{\frac{289}{936}}|\Sigma,D\rangle
-\sqrt{\frac{125}{5616}}|\Sigma_c,\pi\rangle
-\sqrt{\frac{49}{351}}|\Sigma_c,\eta_c\rangle
-\sqrt{\frac{13}{48}}|\Xi_c,\pi\rangle
+\sqrt{\frac{121}{468}}|\Xi_{cc},{\bar D}\rangle
\nonumber \\
(-)~~&
|\mathbf{20_a};\mathbf{6},1\rangle
&=
\phantom{+}\sqrt{\frac{8}{39}}|\Sigma,D\rangle
+\sqrt{\frac{80}{117}}|\Sigma_c,\pi\rangle
-\sqrt{\frac{1}{117}}|\Sigma_c,\eta_c\rangle
-\sqrt{\frac{4}{39}}|\Xi_{cc},{\bar D}\rangle
\nonumber \\
(-)~~&
|\mathbf{20^\prime};\mathbf{6},1\rangle
&=
\phantom{+}\sqrt{\frac{2}{9}}|\Sigma,D\rangle
-\sqrt{\frac{5}{27}}|\Sigma_c,\pi\rangle
-\sqrt{\frac{4}{27}}|\Sigma_c,\eta_c\rangle
+\sqrt{\frac{1}{3}}|\Xi_c,\pi\rangle
-\sqrt{\frac{1}{9}}|\Xi_{cc},{\bar D}\rangle
\nonumber \\
(-)~~&
|\mathbf{60^*};\mathbf{6},1\rangle
&=
\phantom{+}\sqrt{\frac{1}{8}}|\Sigma,D\rangle
-\sqrt{\frac{5}{48}}|\Sigma_c,\pi\rangle
+\sqrt{\frac{1}{3}}|\Sigma_c,\eta_c\rangle
-\sqrt{\frac{3}{16}}|\Xi_c,\pi\rangle
-\sqrt{\frac{1}{4}}|\Xi_{cc},{\bar D}\rangle
\nonumber \\
(+)~~&
|\mathbf{140};\mathbf{6},1\rangle
&=
\phantom{+}\sqrt{\frac{5}{36}}|\Sigma,D\rangle
+\sqrt{\frac{1}{216}}|\Sigma_c,\pi\rangle
+\sqrt{\frac{10}{27}}|\Sigma_c,\eta_c\rangle
+\sqrt{\frac{5}{24}}|\Xi_c,\pi\rangle
+\sqrt{\frac{5}{18}}|\Xi_{cc},{\bar D}\rangle
\end{eqnarray}
\begin{eqnarray}
(-)~~&
|\mathbf{60^*};\mathbf{6^*},-1\rangle
&=
\phantom{+}|\Sigma,{\bar D}\rangle
\end{eqnarray}
\begin{eqnarray}
(+)~~&
|\mathbf{36^*};\mathbf{6^*},2\rangle
&=
\phantom{+}\sqrt{\frac{1}{4}}|\Xi_c,D\rangle
-\sqrt{\frac{3}{4}}|\Xi_{cc},\pi\rangle
\nonumber \\
(+)~~&
|\mathbf{140};\mathbf{6^*},2\rangle
&=
\phantom{+}\sqrt{\frac{3}{4}}|\Xi_c,D\rangle
+\sqrt{\frac{1}{4}}|\Xi_{cc},\pi\rangle
\end{eqnarray}
\begin{eqnarray}
(+)~~&
|\mathbf{20_s};\mathbf{8},0\rangle
&=
\phantom{+}\sqrt{\frac{65}{96}}|(\Sigma,\pi)_s\rangle
-\sqrt{\frac{25}{1248}}|(\Sigma,\pi)_a\rangle
+\sqrt{\frac{1}{156}}|\Sigma,\eta_c\rangle
-\sqrt{\frac{289}{1248}}|\Sigma_c,{\bar D}\rangle
+\sqrt{\frac{27}{416}}|\Xi_c,{\bar D}\rangle
\nonumber \\
(-)~~&
|\mathbf{20_a};\mathbf{8},0\rangle
&=
\phantom{+}\sqrt{\frac{8}{13}}|(\Sigma,\pi)_a\rangle
+\sqrt{\frac{1}{13}}|\Sigma,\eta_c\rangle
-\sqrt{\frac{2}{13}}|\Sigma_c,{\bar D}\rangle
-\sqrt{\frac{2}{13}}|\Xi_c,{\bar D}\rangle
\nonumber \\
(+)~~&
|\mathbf{36^*};\mathbf{8},0\rangle
&=
-\sqrt{\frac{5}{32}}|(\Sigma,\pi)_s\rangle
-\sqrt{\frac{9}{32}}|(\Sigma,\pi)_a\rangle
+\sqrt{\frac{1}{4}}|\Sigma,\eta_c\rangle
-\sqrt{\frac{9}{32}}|\Sigma_c,{\bar D}\rangle
-\sqrt{\frac{1}{32}}|\Xi_c,{\bar D}\rangle
\nonumber \\
(-)~~&
|\mathbf{60^*};\mathbf{8},0\rangle
&=
\phantom{+}\sqrt{\frac{5}{32}}|(\Sigma,\pi)_s\rangle
-\sqrt{\frac{1}{32}}|(\Sigma,\pi)_a\rangle
+\sqrt{\frac{1}{4}}|\Sigma,\eta_c\rangle
+\sqrt{\frac{9}{32}}|\Sigma_c,{\bar D}\rangle
-\sqrt{\frac{9}{32}}|\Xi_c,{\bar D}\rangle
\nonumber \\
(+)~~&
|\mathbf{140};\mathbf{8},0\rangle
&=
-\sqrt{\frac{1}{96}}|(\Sigma,\pi)_s\rangle
+\sqrt{\frac{5}{96}}|(\Sigma,\pi)_a\rangle
+\sqrt{\frac{5}{12}}|\Sigma,\eta_c\rangle
+\sqrt{\frac{5}{96}}|\Sigma_c,{\bar D}\rangle
+\sqrt{\frac{15}{32}}|\Xi_c,{\bar D}\rangle
\end{eqnarray}
\begin{eqnarray}
(+)~~&
|\mathbf{140};\mathbf{8},3\rangle
&=
\phantom{+}|\Xi_{cc},D\rangle
\end{eqnarray}
\begin{eqnarray}
(-)~~&
|\mathbf{20^\prime};\mathbf{10},0\rangle
&=
\phantom{+}\sqrt{\frac{2}{3}}|\Sigma,\pi\rangle
-\sqrt{\frac{1}{3}}|\Sigma_c,{\bar D}\rangle
\nonumber \\
(+)~~&
|\mathbf{140};\mathbf{10},0\rangle
&=
\phantom{+}\sqrt{\frac{1}{3}}|\Sigma,\pi\rangle
+\sqrt{\frac{2}{3}}|\Sigma_c,{\bar D}\rangle
\end{eqnarray}
\begin{eqnarray}
(-)~~&
|\mathbf{60^*};\mathbf{10^*},0\rangle
&=
\phantom{+}|\Sigma,\pi\rangle
\end{eqnarray}
\begin{eqnarray}
(+)~~&
|\mathbf{140};\mathbf{15},-1\rangle
&=
\phantom{+}|\Sigma,{\bar D}\rangle
\end{eqnarray}
\begin{eqnarray}
(-)~~&
|\mathbf{60^*};\mathbf{15},2\rangle
&=
\phantom{+}\sqrt{\frac{1}{2}}|\Sigma_c,D\rangle
-\sqrt{\frac{1}{2}}|\Xi_{cc},\pi\rangle
\nonumber \\
(+)~~&
|\mathbf{140};\mathbf{15},2\rangle
&=
\phantom{+}\sqrt{\frac{1}{2}}|\Sigma_c,D\rangle
+\sqrt{\frac{1}{2}}|\Xi_{cc},\pi\rangle
\end{eqnarray}
\begin{eqnarray}
(+)~~&
|\mathbf{36^*};\mathbf{15^*},1\rangle
&=
\phantom{+}\sqrt{\frac{3}{8}}|\Sigma,D\rangle
-\sqrt{\frac{9}{16}}|\Sigma_c,\pi\rangle
-\sqrt{\frac{1}{16}}|\Xi_c,\pi\rangle
\nonumber \\
(-)~~&
|\mathbf{60^*};\mathbf{15^*},1\rangle
&=
\phantom{+}\sqrt{\frac{1}{4}}|\Sigma,D\rangle
+\sqrt{\frac{3}{8}}|\Sigma_c,\pi\rangle
-\sqrt{\frac{3}{8}}|\Xi_c,\pi\rangle
\nonumber \\
(+)~~&
|\mathbf{140};\mathbf{15^*},1\rangle
&=
\phantom{+}\sqrt{\frac{3}{8}}|\Sigma,D\rangle
+\sqrt{\frac{1}{16}}|\Sigma_c,\pi\rangle
+\sqrt{\frac{9}{16}}|\Xi_c,\pi\rangle
\end{eqnarray}
\begin{eqnarray}
(+)~~&
|\mathbf{140};\mathbf{24^*},1\rangle
&=
\phantom{+}|\Sigma_c,\pi\rangle
\end{eqnarray}
\begin{eqnarray}
(+)~~&
|\mathbf{140};\mathbf{27},0\rangle
&=
\phantom{+}|\Sigma,\pi\rangle
\end{eqnarray}
\subsection{SU(4): $ \mathbf{20^\prime} \otimes \mathbf{15}$} \
\par
\begin{eqnarray}
(-)~~&
|\mathbf{20^\prime};\mathbf{1},3\rangle
&=
\phantom{+}\sqrt{\frac{4}{7}}|\Xi_{cc}^*,D\rangle
-\sqrt{\frac{3}{7}}|\Omega_{ccc},\eta_c\rangle
\nonumber \\
(+)~~&
|\mathbf{120};\mathbf{1},3\rangle
&=
\phantom{+}\sqrt{\frac{3}{7}}|\Xi_{cc}^*,D\rangle
+\sqrt{\frac{4}{7}}|\Omega_{ccc},\eta_c\rangle
\end{eqnarray}
\begin{eqnarray}
(+)~~&
|\mathbf{20};\mathbf{3},2\rangle
&=
\phantom{+}\sqrt{\frac{2}{9}}|\Sigma_c^*,D\rangle
-\sqrt{\frac{8}{27}}|\Xi_{cc}^*,\pi\rangle
-\sqrt{\frac{4}{27}}|\Xi_{cc}^*,\eta_c\rangle
+\sqrt{\frac{1}{3}}|\Omega_{ccc},{\bar D}\rangle
\nonumber \\
(-)~~&
|\mathbf{20^\prime};\mathbf{3},2\rangle
&=
\phantom{+}\sqrt{\frac{32}{63}}|\Sigma_c^*,D\rangle
+\sqrt{\frac{32}{189}}|\Xi_{cc}^*,\pi\rangle
-\sqrt{\frac{25}{189}}|\Xi_{cc}^*,\eta_c\rangle
-\sqrt{\frac{4}{21}}|\Omega_{ccc},{\bar D}\rangle
\nonumber \\
(+)~~&
|\mathbf{120};\mathbf{3},2\rangle
&=
\phantom{+}\sqrt{\frac{3}{14}}|\Sigma_c^*,D\rangle
+\sqrt{\frac{1}{14}}|\Xi_{cc}^*,\pi\rangle
+\sqrt{\frac{4}{7}}|\Xi_{cc}^*,\eta_c\rangle
+\sqrt{\frac{1}{7}}|\Omega_{ccc},{\bar D}\rangle
\nonumber \\
(-)~~&
|\mathbf{140};\mathbf{3},2\rangle
&=
\phantom{+}\sqrt{\frac{1}{18}}|\Sigma_c^*,D\rangle
-\sqrt{\frac{25}{54}}|\Xi_{cc}^*,\pi\rangle
+\sqrt{\frac{4}{27}}|\Xi_{cc}^*,\eta_c\rangle
-\sqrt{\frac{1}{3}}|\Omega_{ccc},{\bar D}\rangle
\end{eqnarray}
\begin{eqnarray}
(+)~~&
|\mathbf{20};\mathbf{3^*},1\rangle
&=
\phantom{+}\sqrt{\frac{2}{3}}|\Sigma_c^*,\pi\rangle
-\sqrt{\frac{1}{3}}|\Xi_{cc}^*,{\bar D}\rangle
\nonumber \\
(-)~~&
|\mathbf{140};\mathbf{3^*},1\rangle
&=
\phantom{+}\sqrt{\frac{1}{3}}|\Sigma_c^*,\pi\rangle
+\sqrt{\frac{2}{3}}|\Xi_{cc}^*,{\bar D}\rangle
\end{eqnarray}
\begin{eqnarray}
(+)~~&
|\mathbf{120};\mathbf{3^*},4\rangle
&=
\phantom{+}|\Omega_{ccc},D\rangle
\end{eqnarray}
\begin{eqnarray}
(+)~~&
|\mathbf{20};\mathbf{6},1\rangle
&=
\phantom{+}\sqrt{\frac{5}{9}}|\Delta,D\rangle
-\sqrt{\frac{5}{27}}|\Sigma_c^*,\pi\rangle
-\sqrt{\frac{4}{27}}|\Sigma_c^*,\eta_c\rangle
+\sqrt{\frac{1}{9}}|\Xi_{cc}^*,{\bar D}\rangle
\nonumber \\
(-)~~&
|\mathbf{20^\prime};\mathbf{6},1\rangle
&=
\phantom{+}\sqrt{\frac{20}{63}}|\Delta,D\rangle
+\sqrt{\frac{80}{189}}|\Sigma_c^*,\pi\rangle
-\sqrt{\frac{1}{189}}|\Sigma_c^*,\eta_c\rangle
-\sqrt{\frac{16}{63}}|\Xi_{cc}^*,{\bar D}\rangle
\nonumber \\
(+)~~&
|\mathbf{120};\mathbf{6},1\rangle
&=
\phantom{+}\sqrt{\frac{1}{14}}|\Delta,D\rangle
+\sqrt{\frac{2}{21}}|\Sigma_c^*,\pi\rangle
+\sqrt{\frac{10}{21}}|\Sigma_c^*,\eta_c\rangle
+\sqrt{\frac{5}{14}}|\Xi_{cc}^*,{\bar D}\rangle
\nonumber \\
(-)~~&
|\mathbf{140};\mathbf{6},1\rangle
&=
\phantom{+}\sqrt{\frac{1}{18}}|\Delta,D\rangle
-\sqrt{\frac{8}{27}}|\Sigma_c^*,\pi\rangle
+\sqrt{\frac{10}{27}}|\Sigma_c^*,\eta_c\rangle
-\sqrt{\frac{5}{18}}|\Xi_{cc}^*,{\bar D}\rangle
\end{eqnarray}
\begin{eqnarray}
(-)~~&
|\mathbf{140};\mathbf{6^*},2\rangle
&=
\phantom{+}|\Xi_{cc}^*,\pi\rangle
\end{eqnarray}
\begin{eqnarray}
(+)~~&
|\mathbf{20};\mathbf{8},0\rangle
&=
\phantom{+}\sqrt{\frac{5}{6}}|\Delta,\pi\rangle
-\sqrt{\frac{1}{6}}|\Sigma_c^*,{\bar D}\rangle
\nonumber \\
(-)~~&
|\mathbf{140};\mathbf{8},0\rangle
&=
\phantom{+}\sqrt{\frac{1}{6}}|\Delta,\pi\rangle
+\sqrt{\frac{5}{6}}|\Sigma_c^*,{\bar D}\rangle
\end{eqnarray}
\begin{eqnarray}
(+)~~&
|\mathbf{120};\mathbf{8},3\rangle
&=
\phantom{+}\sqrt{\frac{3}{4}}|\Xi_{cc}^*,D\rangle
+\sqrt{\frac{1}{4}}|\Omega_{ccc},\pi\rangle
\nonumber \\
(-)~~&
|\mathbf{140};\mathbf{8},3\rangle
&=
\phantom{+}\sqrt{\frac{1}{4}}|\Xi_{cc}^*,D\rangle
-\sqrt{\frac{3}{4}}|\Omega_{ccc},\pi\rangle
\end{eqnarray}
\begin{eqnarray}
(-)~~&
|\mathbf{20^\prime};\mathbf{10},0\rangle
&=
\phantom{+}\sqrt{\frac{16}{21}}|\Delta,\pi\rangle
+\sqrt{\frac{1}{21}}|\Delta,\eta_c\rangle
-\sqrt{\frac{4}{21}}|\Sigma_c^*,{\bar D}\rangle
\nonumber \\
(+)~~&
|\mathbf{120};\mathbf{10},0\rangle
&=
\phantom{+}\sqrt{\frac{1}{14}}|\Delta,\pi\rangle
+\sqrt{\frac{2}{7}}|\Delta,\eta_c\rangle
+\sqrt{\frac{9}{14}}|\Sigma_c^*,{\bar D}\rangle
\nonumber \\
(-)~~&
|\mathbf{140};\mathbf{10},0\rangle
&=
-\sqrt{\frac{1}{6}}|\Delta,\pi\rangle
+\sqrt{\frac{2}{3}}|\Delta,\eta_c\rangle
-\sqrt{\frac{1}{6}}|\Sigma_c^*,{\bar D}\rangle
\end{eqnarray}
\begin{eqnarray}
(-)~~&
|\mathbf{140};\mathbf{15},-1\rangle
&=
\phantom{+}|\Delta,{\bar D}\rangle
\end{eqnarray}
\begin{eqnarray}
(+)~~&
|\mathbf{120};\mathbf{15},2\rangle
&=
\phantom{+}\sqrt{\frac{1}{2}}|\Sigma_c^*,D\rangle
+\sqrt{\frac{1}{2}}|\Xi_{cc}^*,\pi\rangle
\nonumber \\
(-)~~&
|\mathbf{140};\mathbf{15},2\rangle
&=
\phantom{+}\sqrt{\frac{1}{2}}|\Sigma_c^*,D\rangle
-\sqrt{\frac{1}{2}}|\Xi_{cc}^*,\pi\rangle
\end{eqnarray}
\begin{eqnarray}
(+)~~&
|\mathbf{120};\mathbf{15^\prime},-1\rangle
&=
\phantom{+}|\Delta,{\bar D}\rangle
\end{eqnarray}
\begin{eqnarray}
(-)~~&
|\mathbf{140};\mathbf{15^*},1\rangle
&=
\phantom{+}|\Sigma_c^*,\pi\rangle
\end{eqnarray}
\begin{eqnarray}
(+)~~&
|\mathbf{120};\mathbf{24^*},1\rangle
&=
\phantom{+}\sqrt{\frac{1}{4}}|\Delta,D\rangle
+\sqrt{\frac{3}{4}}|\Sigma_c^*,\pi\rangle
\nonumber \\
(-)~~&
|\mathbf{140};\mathbf{24^*},1\rangle
&=
\phantom{+}\sqrt{\frac{3}{4}}|\Delta,D\rangle
-\sqrt{\frac{1}{4}}|\Sigma_c^*,\pi\rangle
\end{eqnarray}
\begin{eqnarray}
(-)~~&
|\mathbf{140};\mathbf{27},0\rangle
&=
\phantom{+}|\Delta,\pi\rangle
\end{eqnarray}
\begin{eqnarray}
(+)~~&
|\mathbf{120};\mathbf{35},0\rangle
&=
\phantom{+}|\Delta,\pi\rangle
\end{eqnarray}
\section{Tables of scalar factors of SU(3)}
\subsection{SU(3): $ \mathbf{3} \otimes \mathbf{3}$} \par
\begin{eqnarray}
(+)~~&
|\mathbf{6};0,-\frac{4}{3}\rangle
&=
\phantom{+}|{\bar D}_s,{\bar D}_s\rangle
\end{eqnarray}
\begin{eqnarray}
(-)~~&
|\mathbf{3^*};0,\frac{2}{3}\rangle
&=
\phantom{+}|{\bar D},{\bar D}\rangle
\end{eqnarray}
\begin{eqnarray}
(-)~~&
|\mathbf{3^*};\frac{1}{2},-\frac{1}{3}\rangle
&=
\phantom{+}|{\bar D},{\bar D}_s\rangle_A
\nonumber \\
(+)~~&
|\mathbf{6};\frac{1}{2},-\frac{1}{3}\rangle
&=
\phantom{+}|{\bar D},{\bar D}_s\rangle_S
\end{eqnarray}
\begin{eqnarray}
(+)~~&
|\mathbf{6};1,\frac{2}{3}\rangle
&=
\phantom{+}|{\bar D},{\bar D}\rangle
\end{eqnarray}
\subsection{SU(3): $ \mathbf{3} \otimes \mathbf{3^*}$} \par
\begin{eqnarray}
(-)~~&
|\mathbf{1};0,0\rangle
&=
\phantom{+}\sqrt{\frac{2}{3}}|{\bar D},D\rangle
+\sqrt{\frac{1}{3}}|{\bar D}_s,D_s\rangle
\nonumber \\
(+)~~&
|\mathbf{8};0,0\rangle
&=
-\sqrt{\frac{1}{3}}|{\bar D},D\rangle
+\sqrt{\frac{2}{3}}|{\bar D}_s,D_s\rangle
\end{eqnarray}
\begin{eqnarray}
(+)~~&
|\mathbf{8};\frac{1}{2},-1\rangle
&=
\phantom{+}|{\bar D}_s,D\rangle
\end{eqnarray}
\begin{eqnarray}
(+)~~&
|\mathbf{8};\frac{1}{2},1\rangle
&=
\phantom{+}|{\bar D},D_s\rangle
\end{eqnarray}
\begin{eqnarray}
(+)~~&
|\mathbf{8};1,0\rangle
&=
\phantom{+}|{\bar D},D\rangle
\end{eqnarray}
\subsection{SU(3): $ \mathbf{3^*} \otimes \mathbf{3^*}$} \par
\begin{eqnarray}
(-)~~&
|\mathbf{3};0,-\frac{2}{3}\rangle
&=
-|D,D\rangle
\end{eqnarray}
\begin{eqnarray}
(+)~~&
|\mathbf{6^*};0,\frac{4}{3}\rangle
&=
\phantom{+}|D_s,D_s\rangle
\end{eqnarray}
\begin{eqnarray}
(-)~~&
|\mathbf{3};\frac{1}{2},\frac{1}{3}\rangle
&=
\phantom{+}|D,D_s\rangle_A
\nonumber \\
(+)~~&
|\mathbf{6^*};\frac{1}{2},\frac{1}{3}\rangle
&=
\phantom{+}|D,D_s\rangle_S
\end{eqnarray}
\begin{eqnarray}
(+)~~&
|\mathbf{6^*};1,-\frac{2}{3}\rangle
&=
\phantom{+}|D,D\rangle
\end{eqnarray}
\subsection{SU(3): $ \mathbf{6} \otimes \mathbf{3}$} \par
\begin{eqnarray}
(+)~~&
|\mathbf{10};0,-2\rangle
&=
\phantom{+}|\Omega_c,{\bar D}_s\rangle
\end{eqnarray}
\begin{eqnarray}
(-)~~&
|\mathbf{8};0,0\rangle
&=
\phantom{+}|\Xi_c^\prime,{\bar D}\rangle
\end{eqnarray}
\begin{eqnarray}
(-)~~&
|\mathbf{8};\frac{1}{2},-1\rangle
&=
\phantom{+}\sqrt{\frac{1}{3}}|\Xi_c^\prime,{\bar D}_s\rangle
-\sqrt{\frac{2}{3}}|\Omega_c,{\bar D}\rangle
\nonumber \\
(+)~~&
|\mathbf{10};\frac{1}{2},-1\rangle
&=
\phantom{+}\sqrt{\frac{2}{3}}|\Xi_c^\prime,{\bar D}_s\rangle
+\sqrt{\frac{1}{3}}|\Omega_c,{\bar D}\rangle
\end{eqnarray}
\begin{eqnarray}
(-)~~&
|\mathbf{8};\frac{1}{2},1\rangle
&=
\phantom{+}|\Sigma_c,{\bar D}\rangle
\end{eqnarray}
\begin{eqnarray}
(-)~~&
|\mathbf{8};1,0\rangle
&=
\phantom{+}\sqrt{\frac{2}{3}}|\Sigma_c,{\bar D}_s\rangle
-\sqrt{\frac{1}{3}}|\Xi_c^\prime,{\bar D}\rangle
\nonumber \\
(+)~~&
|\mathbf{10};1,0\rangle
&=
\phantom{+}\sqrt{\frac{1}{3}}|\Sigma_c,{\bar D}_s\rangle
+\sqrt{\frac{2}{3}}|\Xi_c^\prime,{\bar D}\rangle
\end{eqnarray}
\begin{eqnarray}
(+)~~&
|\mathbf{10};\frac{3}{2},1\rangle
&=
\phantom{+}|\Sigma_c,{\bar D}\rangle
\end{eqnarray}
\subsection{SU(3): $ \mathbf{6} \otimes \mathbf{3^*}$} \par
\begin{eqnarray}
(-)~~&
|\mathbf{3};0,-\frac{2}{3}\rangle
&=
\phantom{+}\sqrt{\frac{1}{2}}|\Xi_c^\prime,D\rangle
+\sqrt{\frac{1}{2}}|\Omega_c,D_s\rangle
\nonumber \\
(+)~~&
|\mathbf{15};0,-\frac{2}{3}\rangle
&=
-\sqrt{\frac{1}{2}}|\Xi_c^\prime,D\rangle
+\sqrt{\frac{1}{2}}|\Omega_c,D_s\rangle
\end{eqnarray}
\begin{eqnarray}
(+)~~&
|\mathbf{15};\frac{1}{2},-\frac{5}{3}\rangle
&=
\phantom{+}|\Omega_c,D\rangle
\end{eqnarray}
\begin{eqnarray}
(-)~~&
|\mathbf{3};\frac{1}{2},\frac{1}{3}\rangle
&=
\phantom{+}\sqrt{\frac{3}{4}}|\Sigma_c,D\rangle
+\sqrt{\frac{1}{4}}|\Xi_c^\prime,D_s\rangle
\nonumber \\
(+)~~&
|\mathbf{15};\frac{1}{2},\frac{1}{3}\rangle
&=
-\sqrt{\frac{1}{4}}|\Sigma_c,D\rangle
+\sqrt{\frac{3}{4}}|\Xi_c^\prime,D_s\rangle
\end{eqnarray}
\begin{eqnarray}
(+)~~&
|\mathbf{15};1,-\frac{2}{3}\rangle
&=
\phantom{+}|\Xi_c^\prime,D\rangle
\end{eqnarray}
\begin{eqnarray}
(+)~~&
|\mathbf{15};1,\frac{4}{3}\rangle
&=
\phantom{+}|\Sigma_c,D_s\rangle
\end{eqnarray}
\begin{eqnarray}
(+)~~&
|\mathbf{15};\frac{3}{2},\frac{1}{3}\rangle
&=
\phantom{+}|\Sigma_c,D\rangle
\end{eqnarray}
\subsection{SU(3): $ \mathbf{6} \otimes \mathbf{8}$} \par
\begin{eqnarray}
(-)~~&
|\mathbf{6};0,-\frac{4}{3}\rangle
&=
\phantom{+}\sqrt{\frac{3}{5}}|\Xi_c^\prime,{\bar K}\rangle
+\sqrt{\frac{2}{5}}|\Omega_c,\eta\rangle
\nonumber \\
(+)~~&
|\mathbf{24^*};0,-\frac{4}{3}\rangle
&=
-\sqrt{\frac{2}{5}}|\Xi_c^\prime,{\bar K}\rangle
+\sqrt{\frac{3}{5}}|\Omega_c,\eta\rangle
\end{eqnarray}
\begin{eqnarray}
(+)~~&
|\mathbf{3^*};0,\frac{2}{3}\rangle
&=
\phantom{+}\sqrt{\frac{3}{4}}|\Sigma_c,\pi\rangle
+\sqrt{\frac{1}{4}}|\Xi_c^\prime,K\rangle
\nonumber \\
(-)~~&
|\mathbf{15^*};0,\frac{2}{3}\rangle
&=
-\sqrt{\frac{1}{4}}|\Sigma_c,\pi\rangle
+\sqrt{\frac{3}{4}}|\Xi_c^\prime,K\rangle
\end{eqnarray}
\begin{eqnarray}
(+)~~&
|\mathbf{24^*};\frac{1}{2},-\frac{7}{3}\rangle
&=
\phantom{+}|\Omega_c,{\bar K}\rangle
\end{eqnarray}
\begin{eqnarray}
(+)~~&
|\mathbf{3^*};\frac{1}{2},-\frac{1}{3}\rangle
&=
\phantom{+}\sqrt{\frac{3}{8}}|\Sigma_c,{\bar K}\rangle
-\sqrt{\frac{3}{16}}|\Xi_c^\prime,\pi\rangle
+\sqrt{\frac{3}{16}}|\Xi_c^\prime,\eta\rangle
-\sqrt{\frac{1}{4}}|\Omega_c,K\rangle
\nonumber \\
(-)~~&
|\mathbf{6};\frac{1}{2},-\frac{1}{3}\rangle
&=
\phantom{+}\sqrt{\frac{9}{20}}|\Sigma_c,{\bar K}\rangle
+\sqrt{\frac{9}{40}}|\Xi_c^\prime,\pi\rangle
+\sqrt{\frac{1}{40}}|\Xi_c^\prime,\eta\rangle
+\sqrt{\frac{3}{10}}|\Omega_c,K\rangle
\nonumber \\
(-)~~&
|\mathbf{15^*};\frac{1}{2},-\frac{1}{3}\rangle
&=
-\sqrt{\frac{1}{24}}|\Sigma_c,{\bar K}\rangle
+\sqrt{\frac{25}{48}}|\Xi_c^\prime,\pi\rangle
+\sqrt{\frac{3}{16}}|\Xi_c^\prime,\eta\rangle
-\sqrt{\frac{1}{4}}|\Omega_c,K\rangle
\nonumber \\
(+)~~&
|\mathbf{24^*};\frac{1}{2},-\frac{1}{3}\rangle
&=
-\sqrt{\frac{2}{15}}|\Sigma_c,{\bar K}\rangle
-\sqrt{\frac{1}{15}}|\Xi_c^\prime,\pi\rangle
+\sqrt{\frac{3}{5}}|\Xi_c^\prime,\eta\rangle
+\sqrt{\frac{1}{5}}|\Omega_c,K\rangle
\end{eqnarray}
\begin{eqnarray}
(-)~~&
|\mathbf{15^*};\frac{1}{2},\frac{5}{3}\rangle
&=
\phantom{+}|\Sigma_c,K\rangle
\end{eqnarray}
\begin{eqnarray}
(-)~~&
|\mathbf{15^*};1,-\frac{4}{3}\rangle
&=
\phantom{+}\sqrt{\frac{1}{3}}|\Xi_c^\prime,{\bar K}\rangle
-\sqrt{\frac{2}{3}}|\Omega_c,\pi\rangle
\nonumber \\
(+)~~&
|\mathbf{24^*};1,-\frac{4}{3}\rangle
&=
\phantom{+}\sqrt{\frac{2}{3}}|\Xi_c^\prime,{\bar K}\rangle
+\sqrt{\frac{1}{3}}|\Omega_c,\pi\rangle
\end{eqnarray}
\begin{eqnarray}
(-)~~&
|\mathbf{6};1,\frac{2}{3}\rangle
&=
\phantom{+}\sqrt{\frac{3}{5}}|\Sigma_c,\pi\rangle
-\sqrt{\frac{1}{10}}|\Sigma_c,\eta\rangle
+\sqrt{\frac{3}{10}}|\Xi_c^\prime,K\rangle
\nonumber \\
(-)~~&
|\mathbf{15^*};1,\frac{2}{3}\rangle
&=
\phantom{+}\sqrt{\frac{1}{3}}|\Sigma_c,\pi\rangle
+\sqrt{\frac{1}{2}}|\Sigma_c,\eta\rangle
-\sqrt{\frac{1}{6}}|\Xi_c^\prime,K\rangle
\nonumber \\
(+)~~&
|\mathbf{24^*};1,\frac{2}{3}\rangle
&=
-\sqrt{\frac{1}{15}}|\Sigma_c,\pi\rangle
+\sqrt{\frac{2}{5}}|\Sigma_c,\eta\rangle
+\sqrt{\frac{8}{15}}|\Xi_c^\prime,K\rangle
\end{eqnarray}
\begin{eqnarray}
(-)~~&
|\mathbf{15^*};\frac{3}{2},-\frac{1}{3}\rangle
&=
\phantom{+}\sqrt{\frac{2}{3}}|\Sigma_c,{\bar K}\rangle
-\sqrt{\frac{1}{3}}|\Xi_c^\prime,\pi\rangle
\nonumber \\
(+)~~&
|\mathbf{24^*};\frac{3}{2},-\frac{1}{3}\rangle
&=
\phantom{+}\sqrt{\frac{1}{3}}|\Sigma_c,{\bar K}\rangle
+\sqrt{\frac{2}{3}}|\Xi_c^\prime,\pi\rangle
\end{eqnarray}
\begin{eqnarray}
(+)~~&
|\mathbf{24^*};\frac{3}{2},\frac{5}{3}\rangle
&=
\phantom{+}|\Sigma_c,K\rangle
\end{eqnarray}
\begin{eqnarray}
(+)~~&
|\mathbf{24^*};2,\frac{2}{3}\rangle
&=
\phantom{+}|\Sigma_c,\pi\rangle
\end{eqnarray}
\subsection{SU(3): $ \mathbf{8} \otimes \mathbf{3}$} \par
\begin{eqnarray}
(-)~~&
|\mathbf{3};0,-\frac{2}{3}\rangle
&=
\phantom{+}\sqrt{\frac{3}{4}}|{\bar K},{\bar D}\rangle
-\sqrt{\frac{1}{4}}|\eta,{\bar D}_s\rangle
\nonumber \\
(+)~~&
|\mathbf{15};0,-\frac{2}{3}\rangle
&=
\phantom{+}\sqrt{\frac{1}{4}}|{\bar K},{\bar D}\rangle
+\sqrt{\frac{3}{4}}|\eta,{\bar D}_s\rangle
\end{eqnarray}
\begin{eqnarray}
(-)~~&
|\mathbf{6^*};0,\frac{4}{3}\rangle
&=
\phantom{+}|K,{\bar D}\rangle
\end{eqnarray}
\begin{eqnarray}
(+)~~&
|\mathbf{15};\frac{1}{2},-\frac{5}{3}\rangle
&=
\phantom{+}|{\bar K},{\bar D}_s\rangle
\end{eqnarray}
\begin{eqnarray}
(-)~~&
|\mathbf{3};\frac{1}{2},\frac{1}{3}\rangle
&=
\phantom{+}\sqrt{\frac{9}{16}}|\pi,{\bar D}\rangle
-\sqrt{\frac{3}{8}}|K,{\bar D}_s\rangle
+\sqrt{\frac{1}{16}}|\eta,{\bar D}\rangle
\nonumber \\
(-)~~&
|\mathbf{6^*};\frac{1}{2},\frac{1}{3}\rangle
&=
\phantom{+}\sqrt{\frac{3}{8}}|\pi,{\bar D}\rangle
+\sqrt{\frac{1}{4}}|K,{\bar D}_s\rangle
-\sqrt{\frac{3}{8}}|\eta,{\bar D}\rangle
\nonumber \\
(+)~~&
|\mathbf{15};\frac{1}{2},\frac{1}{3}\rangle
&=
\phantom{+}\sqrt{\frac{1}{16}}|\pi,{\bar D}\rangle
+\sqrt{\frac{3}{8}}|K,{\bar D}_s\rangle
+\sqrt{\frac{9}{16}}|\eta,{\bar D}\rangle
\end{eqnarray}
\begin{eqnarray}
(-)~~&
|\mathbf{6^*};1,-\frac{2}{3}\rangle
&=
\phantom{+}\sqrt{\frac{1}{2}}|\pi,{\bar D}_s\rangle
-\sqrt{\frac{1}{2}}|{\bar K},{\bar D}\rangle
\nonumber \\
(+)~~&
|\mathbf{15};1,-\frac{2}{3}\rangle
&=
\phantom{+}\sqrt{\frac{1}{2}}|\pi,{\bar D}_s\rangle
+\sqrt{\frac{1}{2}}|{\bar K},{\bar D}\rangle
\end{eqnarray}
\begin{eqnarray}
(+)~~&
|\mathbf{15};1,\frac{4}{3}\rangle
&=
\phantom{+}|K,{\bar D}\rangle
\end{eqnarray}
\begin{eqnarray}
(+)~~&
|\mathbf{15};\frac{3}{2},\frac{1}{3}\rangle
&=
\phantom{+}|\pi,{\bar D}\rangle
\end{eqnarray}
\subsection{SU(3): $ \mathbf{8} \otimes \mathbf{3^*}$} \par
\begin{eqnarray}
(-)~~&
|\mathbf{6};0,-\frac{4}{3}\rangle
&=
-|{\bar K},D\rangle
\end{eqnarray}
\begin{eqnarray}
(-)~~&
|\mathbf{3^*};0,\frac{2}{3}\rangle
&=
\phantom{+}\sqrt{\frac{3}{4}}|K,D\rangle
+\sqrt{\frac{1}{4}}|\eta,D_s\rangle
\nonumber \\
(+)~~&
|\mathbf{15^*};0,\frac{2}{3}\rangle
&=
-\sqrt{\frac{1}{4}}|K,D\rangle
+\sqrt{\frac{3}{4}}|\eta,D_s\rangle
\end{eqnarray}
\begin{eqnarray}
(-)~~&
|\mathbf{3^*};\frac{1}{2},-\frac{1}{3}\rangle
&=
\phantom{+}\sqrt{\frac{9}{16}}|\pi,D\rangle
+\sqrt{\frac{3}{8}}|{\bar K},D_s\rangle
-\sqrt{\frac{1}{16}}|\eta,D\rangle
\nonumber \\
(-)~~&
|\mathbf{6};\frac{1}{2},-\frac{1}{3}\rangle
&=
-\sqrt{\frac{3}{8}}|\pi,D\rangle
+\sqrt{\frac{1}{4}}|{\bar K},D_s\rangle
-\sqrt{\frac{3}{8}}|\eta,D\rangle
\nonumber \\
(+)~~&
|\mathbf{15^*};\frac{1}{2},-\frac{1}{3}\rangle
&=
-\sqrt{\frac{1}{16}}|\pi,D\rangle
+\sqrt{\frac{3}{8}}|{\bar K},D_s\rangle
+\sqrt{\frac{9}{16}}|\eta,D\rangle
\end{eqnarray}
\begin{eqnarray}
(+)~~&
|\mathbf{15^*};\frac{1}{2},\frac{5}{3}\rangle
&=
\phantom{+}|K,D_s\rangle
\end{eqnarray}
\begin{eqnarray}
(+)~~&
|\mathbf{15^*};1,-\frac{4}{3}\rangle
&=
\phantom{+}|{\bar K},D\rangle
\end{eqnarray}
\begin{eqnarray}
(-)~~&
|\mathbf{6};1,\frac{2}{3}\rangle
&=
\phantom{+}\sqrt{\frac{1}{2}}|\pi,D_s\rangle
-\sqrt{\frac{1}{2}}|K,D\rangle
\nonumber \\
(+)~~&
|\mathbf{15^*};1,\frac{2}{3}\rangle
&=
\phantom{+}\sqrt{\frac{1}{2}}|\pi,D_s\rangle
+\sqrt{\frac{1}{2}}|K,D\rangle
\end{eqnarray}
\begin{eqnarray}
(+)~~&
|\mathbf{15^*};\frac{3}{2},-\frac{1}{3}\rangle
&=
\phantom{+}|\pi,D\rangle
\end{eqnarray}
\subsection{SU(3): $ \mathbf{8} \otimes \mathbf{8}$} \par
\begin{eqnarray}
(-)~~&
|\mathbf{10};0,-2\rangle
&=
-|{\bar K},{\bar K}\rangle
\end{eqnarray}
\begin{eqnarray}
(+)~~&
|\mathbf{1};0,0\rangle
&=
\phantom{+}\sqrt{\frac{3}{8}}|\pi,\pi\rangle
-\sqrt{\frac{1}{2}}|K,{\bar K}\rangle_A
-\sqrt{\frac{1}{8}}|\eta,\eta\rangle
\nonumber \\
(+)~~&
|\mathbf{8_s};0,0\rangle
&=
-\sqrt{\frac{3}{5}}|\pi,\pi\rangle
-\sqrt{\frac{1}{5}}|K,{\bar K}\rangle_A
-\sqrt{\frac{1}{5}}|\eta,\eta\rangle
\nonumber \\
(-)~~&
|\mathbf{8_a};0,0\rangle
&=
\phantom{+}|K,{\bar K}\rangle_S
\nonumber \\
(+)~~&
|\mathbf{27};0,0\rangle
&=
-\sqrt{\frac{1}{40}}|\pi,\pi\rangle
-\sqrt{\frac{3}{10}}|K,{\bar K}\rangle_A
+\sqrt{\frac{27}{40}}|\eta,\eta\rangle
\end{eqnarray}
\begin{eqnarray}
(-)~~&
|\mathbf{10^*};0,2\rangle
&=
\phantom{+}|K,K\rangle
\end{eqnarray}
\begin{eqnarray}
(+)~~&
|\mathbf{8_s};\frac{1}{2},-1\rangle
&=
-\sqrt{\frac{9}{10}}|\pi,{\bar K}\rangle_A
-\sqrt{\frac{1}{10}}|{\bar K},\eta\rangle_S
\nonumber \\
(-)~~&
|\mathbf{8_a};\frac{1}{2},-1\rangle
&=
\phantom{+}\sqrt{\frac{1}{2}}|\pi,{\bar K}\rangle_S
+\sqrt{\frac{1}{2}}|{\bar K},\eta\rangle_A
\nonumber \\
(-)~~&
|\mathbf{10};\frac{1}{2},-1\rangle
&=
-\sqrt{\frac{1}{2}}|\pi,{\bar K}\rangle_S
+\sqrt{\frac{1}{2}}|{\bar K},\eta\rangle_A
\nonumber \\
(+)~~&
|\mathbf{27};\frac{1}{2},-1\rangle
&=
-\sqrt{\frac{1}{10}}|\pi,{\bar K}\rangle_A
+\sqrt{\frac{9}{10}}|{\bar K},\eta\rangle_S
\end{eqnarray}
\begin{eqnarray}
(+)~~&
|\mathbf{8_s};\frac{1}{2},1\rangle
&=
\phantom{+}\sqrt{\frac{9}{10}}|\pi,K\rangle_A
-\sqrt{\frac{1}{10}}|K,\eta\rangle_S
\nonumber \\
(-)~~&
|\mathbf{8_a};\frac{1}{2},1\rangle
&=
\phantom{+}\sqrt{\frac{1}{2}}|\pi,K\rangle_S
-\sqrt{\frac{1}{2}}|K,\eta\rangle_A
\nonumber \\
(-)~~&
|\mathbf{10^*};\frac{1}{2},1\rangle
&=
\phantom{+}\sqrt{\frac{1}{2}}|\pi,K\rangle_S
+\sqrt{\frac{1}{2}}|K,\eta\rangle_A
\nonumber \\
(+)~~&
|\mathbf{27};\frac{1}{2},1\rangle
&=
\phantom{+}\sqrt{\frac{1}{10}}|\pi,K\rangle_A
+\sqrt{\frac{9}{10}}|K,\eta\rangle_S
\end{eqnarray}
\begin{eqnarray}
(+)~~&
|\mathbf{27};1,-2\rangle
&=
\phantom{+}|{\bar K},{\bar K}\rangle
\end{eqnarray}
\begin{eqnarray}
(+)~~&
|\mathbf{8_s};1,0\rangle
&=
\phantom{+}\sqrt{\frac{2}{5}}|\pi,\eta\rangle_S
-\sqrt{\frac{3}{5}}|K,{\bar K}\rangle_S
\nonumber \\
(-)~~&
|\mathbf{8_a};1,0\rangle
&=
\phantom{+}\sqrt{\frac{2}{3}}|\pi,\pi\rangle
-\sqrt{\frac{1}{3}}|K,{\bar K}\rangle_A
\nonumber \\
(-)~~&
|\mathbf{10};1,0\rangle
&=
-\sqrt{\frac{1}{6}}|\pi,\pi\rangle
+\sqrt{\frac{1}{2}}|\pi,\eta\rangle_A
-\sqrt{\frac{1}{3}}|K,{\bar K}\rangle_A
\nonumber \\
(-)~~&
|\mathbf{10^*};1,0\rangle
&=
\phantom{+}\sqrt{\frac{1}{6}}|\pi,\pi\rangle
+\sqrt{\frac{1}{2}}|\pi,\eta\rangle_A
+\sqrt{\frac{1}{3}}|K,{\bar K}\rangle_A
\nonumber \\
(+)~~&
|\mathbf{27};1,0\rangle
&=
\phantom{+}\sqrt{\frac{3}{5}}|\pi,\eta\rangle_S
+\sqrt{\frac{2}{5}}|K,{\bar K}\rangle_S
\end{eqnarray}
\begin{eqnarray}
(+)~~&
|\mathbf{27};1,2\rangle
&=
\phantom{+}|K,K\rangle
\end{eqnarray}
\begin{eqnarray}
(-)~~&
|\mathbf{10^*};\frac{3}{2},-1\rangle
&=
\phantom{+}|\pi,{\bar K}\rangle_A
\nonumber \\
(+)~~&
|\mathbf{27};\frac{3}{2},-1\rangle
&=
\phantom{+}|\pi,{\bar K}\rangle_S
\end{eqnarray}
\begin{eqnarray}
(-)~~&
|\mathbf{10};\frac{3}{2},1\rangle
&=
\phantom{+}|\pi,K\rangle_A
\nonumber \\
(+)~~&
|\mathbf{27};\frac{3}{2},1\rangle
&=
\phantom{+}|\pi,K\rangle_S
\end{eqnarray}
\begin{eqnarray}
(+)~~&
|\mathbf{27};2,0\rangle
&=
\phantom{+}|\pi,\pi\rangle
\end{eqnarray}
\subsection{SU(3): $ \mathbf{10} \otimes \mathbf{3}$} \par
\begin{eqnarray}
(+)~~&
|\mathbf{15^\prime};0,-\frac{8}{3}\rangle
&=
\phantom{+}|\Omega,{\bar D}_s\rangle
\end{eqnarray}
\begin{eqnarray}
(-)~~&
|\mathbf{15};0,-\frac{2}{3}\rangle
&=
\phantom{+}|\Xi^*,{\bar D}\rangle
\end{eqnarray}
\begin{eqnarray}
(-)~~&
|\mathbf{15};\frac{1}{2},-\frac{5}{3}\rangle
&=
\phantom{+}\sqrt{\frac{1}{4}}|\Xi^*,{\bar D}_s\rangle
-\sqrt{\frac{3}{4}}|\Omega,{\bar D}\rangle
\nonumber \\
(+)~~&
|\mathbf{15^\prime};\frac{1}{2},-\frac{5}{3}\rangle
&=
\phantom{+}\sqrt{\frac{3}{4}}|\Xi^*,{\bar D}_s\rangle
+\sqrt{\frac{1}{4}}|\Omega,{\bar D}\rangle
\end{eqnarray}
\begin{eqnarray}
(-)~~&
|\mathbf{15};\frac{1}{2},\frac{1}{3}\rangle
&=
\phantom{+}|\Sigma^*,{\bar D}\rangle
\end{eqnarray}
\begin{eqnarray}
(-)~~&
|\mathbf{15};1,-\frac{2}{3}\rangle
&=
\phantom{+}\sqrt{\frac{1}{2}}|\Sigma^*,{\bar D}_s\rangle
-\sqrt{\frac{1}{2}}|\Xi^*,{\bar D}\rangle
\nonumber \\
(+)~~&
|\mathbf{15^\prime};1,-\frac{2}{3}\rangle
&=
\phantom{+}\sqrt{\frac{1}{2}}|\Sigma^*,{\bar D}_s\rangle
+\sqrt{\frac{1}{2}}|\Xi^*,{\bar D}\rangle
\end{eqnarray}
\begin{eqnarray}
(-)~~&
|\mathbf{15};1,\frac{4}{3}\rangle
&=
\phantom{+}|\Delta,{\bar D}\rangle
\end{eqnarray}
\begin{eqnarray}
(-)~~&
|\mathbf{15};\frac{3}{2},\frac{1}{3}\rangle
&=
\phantom{+}\sqrt{\frac{3}{4}}|\Delta,{\bar D}_s\rangle
-\sqrt{\frac{1}{4}}|\Sigma^*,{\bar D}\rangle
\nonumber \\
(+)~~&
|\mathbf{15^\prime};\frac{3}{2},\frac{1}{3}\rangle
&=
\phantom{+}\sqrt{\frac{1}{4}}|\Delta,{\bar D}_s\rangle
+\sqrt{\frac{3}{4}}|\Sigma^*,{\bar D}\rangle
\end{eqnarray}
\begin{eqnarray}
(+)~~&
|\mathbf{15^\prime};2,\frac{4}{3}\rangle
&=
\phantom{+}|\Delta,{\bar D}\rangle
\end{eqnarray}
\subsection{SU(3): $ \mathbf{10} \otimes \mathbf{3^*}$} \par
\begin{eqnarray}
(-)~~&
|\mathbf{6};0,-\frac{4}{3}\rangle
&=
\phantom{+}\sqrt{\frac{2}{5}}|\Xi^*,D\rangle
+\sqrt{\frac{3}{5}}|\Omega,D_s\rangle
\nonumber \\
(+)~~&
|\mathbf{24^*};0,-\frac{4}{3}\rangle
&=
-\sqrt{\frac{3}{5}}|\Xi^*,D\rangle
+\sqrt{\frac{2}{5}}|\Omega,D_s\rangle
\end{eqnarray}
\begin{eqnarray}
(+)~~&
|\mathbf{24^*};\frac{1}{2},-\frac{7}{3}\rangle
&=
\phantom{+}|\Omega,D\rangle
\end{eqnarray}
\begin{eqnarray}
(-)~~&
|\mathbf{6};\frac{1}{2},-\frac{1}{3}\rangle
&=
\phantom{+}\sqrt{\frac{3}{5}}|\Sigma^*,D\rangle
+\sqrt{\frac{2}{5}}|\Xi^*,D_s\rangle
\nonumber \\
(+)~~&
|\mathbf{24^*};\frac{1}{2},-\frac{1}{3}\rangle
&=
-\sqrt{\frac{2}{5}}|\Sigma^*,D\rangle
+\sqrt{\frac{3}{5}}|\Xi^*,D_s\rangle
\end{eqnarray}
\begin{eqnarray}
(+)~~&
|\mathbf{24^*};1,-\frac{4}{3}\rangle
&=
\phantom{+}|\Xi^*,D\rangle
\end{eqnarray}
\begin{eqnarray}
(-)~~&
|\mathbf{6};1,\frac{2}{3}\rangle
&=
\phantom{+}\sqrt{\frac{4}{5}}|\Delta,D\rangle
+\sqrt{\frac{1}{5}}|\Sigma^*,D_s\rangle
\nonumber \\
(+)~~&
|\mathbf{24^*};1,\frac{2}{3}\rangle
&=
-\sqrt{\frac{1}{5}}|\Delta,D\rangle
+\sqrt{\frac{4}{5}}|\Sigma^*,D_s\rangle
\end{eqnarray}
\begin{eqnarray}
(+)~~&
|\mathbf{24^*};\frac{3}{2},-\frac{1}{3}\rangle
&=
\phantom{+}|\Sigma^*,D\rangle
\end{eqnarray}
\begin{eqnarray}
(+)~~&
|\mathbf{24^*};\frac{3}{2},\frac{5}{3}\rangle
&=
\phantom{+}|\Delta,D_s\rangle
\end{eqnarray}
\begin{eqnarray}
(+)~~&
|\mathbf{24^*};2,\frac{2}{3}\rangle
&=
\phantom{+}|\Delta,D\rangle
\end{eqnarray}
\subsection{SU(3): $ \mathbf{10} \otimes \mathbf{8}$} \par
\begin{eqnarray}
(-)~~&
|\mathbf{10};0,-2\rangle
&=
\phantom{+}\sqrt{\frac{1}{2}}|\Xi^*,{\bar K}\rangle
+\sqrt{\frac{1}{2}}|\Omega,\eta\rangle
\nonumber \\
(+)~~&
|\mathbf{35};0,-2\rangle
&=
-\sqrt{\frac{1}{2}}|\Xi^*,{\bar K}\rangle
+\sqrt{\frac{1}{2}}|\Omega,\eta\rangle
\end{eqnarray}
\begin{eqnarray}
(+)~~&
|\mathbf{8};0,0\rangle
&=
\phantom{+}\sqrt{\frac{3}{5}}|\Sigma^*,\pi\rangle
+\sqrt{\frac{2}{5}}|\Xi^*,K\rangle
\nonumber \\
(-)~~&
|\mathbf{27};0,0\rangle
&=
-\sqrt{\frac{2}{5}}|\Sigma^*,\pi\rangle
+\sqrt{\frac{3}{5}}|\Xi^*,K\rangle
\end{eqnarray}
\begin{eqnarray}
(+)~~&
|\mathbf{35};\frac{1}{2},-3\rangle
&=
\phantom{+}|\Omega,{\bar K}\rangle
\end{eqnarray}
\begin{eqnarray}
(+)~~&
|\mathbf{8};\frac{1}{2},-1\rangle
&=
\phantom{+}\sqrt{\frac{1}{5}}|\Sigma^*,{\bar K}\rangle
-\sqrt{\frac{1}{5}}|\Xi^*,\pi\rangle
+\sqrt{\frac{1}{5}}|\Xi^*,\eta\rangle
-\sqrt{\frac{2}{5}}|\Omega,K\rangle
\nonumber \\
(-)~~&
|\mathbf{10};\frac{1}{2},-1\rangle
&=
\phantom{+}\sqrt{\frac{1}{2}}|\Sigma^*,{\bar K}\rangle
+\sqrt{\frac{1}{8}}|\Xi^*,\pi\rangle
+\sqrt{\frac{1}{8}}|\Xi^*,\eta\rangle
+\sqrt{\frac{1}{4}}|\Omega,K\rangle
\nonumber \\
(-)~~&
|\mathbf{27};\frac{1}{2},-1\rangle
&=
-\sqrt{\frac{1}{20}}|\Sigma^*,{\bar K}\rangle
+\sqrt{\frac{49}{80}}|\Xi^*,\pi\rangle
+\sqrt{\frac{9}{80}}|\Xi^*,\eta\rangle
-\sqrt{\frac{9}{40}}|\Omega,K\rangle
\nonumber \\
(+)~~&
|\mathbf{35};\frac{1}{2},-1\rangle
&=
-\sqrt{\frac{1}{4}}|\Sigma^*,{\bar K}\rangle
-\sqrt{\frac{1}{16}}|\Xi^*,\pi\rangle
+\sqrt{\frac{9}{16}}|\Xi^*,\eta\rangle
+\sqrt{\frac{1}{8}}|\Omega,K\rangle
\end{eqnarray}
\begin{eqnarray}
(+)~~&
|\mathbf{8};\frac{1}{2},1\rangle
&=
\phantom{+}\sqrt{\frac{4}{5}}|\Delta,\pi\rangle
+\sqrt{\frac{1}{5}}|\Sigma^*,K\rangle
\nonumber \\
(-)~~&
|\mathbf{27};\frac{1}{2},1\rangle
&=
-\sqrt{\frac{1}{5}}|\Delta,\pi\rangle
+\sqrt{\frac{4}{5}}|\Sigma^*,K\rangle
\end{eqnarray}
\begin{eqnarray}
(-)~~&
|\mathbf{27};1,-2\rangle
&=
\phantom{+}\sqrt{\frac{1}{4}}|\Xi^*,{\bar K}\rangle
-\sqrt{\frac{3}{4}}|\Omega,\pi\rangle
\nonumber \\
(+)~~&
|\mathbf{35};1,-2\rangle
&=
\phantom{+}\sqrt{\frac{3}{4}}|\Xi^*,{\bar K}\rangle
+\sqrt{\frac{1}{4}}|\Omega,\pi\rangle
\end{eqnarray}
\begin{eqnarray}
(+)~~&
|\mathbf{8};1,0\rangle
&=
\phantom{+}\sqrt{\frac{8}{15}}|\Delta,{\bar K}\rangle
-\sqrt{\frac{2}{15}}|\Sigma^*,\pi\rangle
+\sqrt{\frac{1}{5}}|\Sigma^*,\eta\rangle
-\sqrt{\frac{2}{15}}|\Xi^*,K\rangle
\nonumber \\
(-)~~&
|\mathbf{10};1,0\rangle
&=
\phantom{+}\sqrt{\frac{1}{3}}|\Delta,{\bar K}\rangle
+\sqrt{\frac{1}{3}}|\Sigma^*,\pi\rangle
+\sqrt{\frac{1}{3}}|\Xi^*,K\rangle
\nonumber \\
(-)~~&
|\mathbf{27};1,0\rangle
&=
-\sqrt{\frac{1}{20}}|\Delta,{\bar K}\rangle
+\sqrt{\frac{9}{20}}|\Sigma^*,\pi\rangle
+\sqrt{\frac{3}{10}}|\Sigma^*,\eta\rangle
-\sqrt{\frac{1}{5}}|\Xi^*,K\rangle
\nonumber \\
(+)~~&
|\mathbf{35};1,0\rangle
&=
-\sqrt{\frac{1}{12}}|\Delta,{\bar K}\rangle
-\sqrt{\frac{1}{12}}|\Sigma^*,\pi\rangle
+\sqrt{\frac{1}{2}}|\Sigma^*,\eta\rangle
+\sqrt{\frac{1}{3}}|\Xi^*,K\rangle
\end{eqnarray}
\begin{eqnarray}
(-)~~&
|\mathbf{27};1,2\rangle
&=
\phantom{+}|\Delta,K\rangle
\end{eqnarray}
\begin{eqnarray}
(-)~~&
|\mathbf{27};\frac{3}{2},-1\rangle
&=
\phantom{+}\sqrt{\frac{1}{2}}|\Sigma^*,{\bar K}\rangle
-\sqrt{\frac{1}{2}}|\Xi^*,\pi\rangle
\nonumber \\
(+)~~&
|\mathbf{35};\frac{3}{2},-1\rangle
&=
\phantom{+}\sqrt{\frac{1}{2}}|\Sigma^*,{\bar K}\rangle
+\sqrt{\frac{1}{2}}|\Xi^*,\pi\rangle
\end{eqnarray}
\begin{eqnarray}
(-)~~&
|\mathbf{10};\frac{3}{2},1\rangle
&=
\phantom{+}\sqrt{\frac{5}{8}}|\Delta,\pi\rangle
-\sqrt{\frac{1}{8}}|\Delta,\eta\rangle
+\sqrt{\frac{1}{4}}|\Sigma^*,K\rangle
\nonumber \\
(-)~~&
|\mathbf{27};\frac{3}{2},1\rangle
&=
\phantom{+}\sqrt{\frac{5}{16}}|\Delta,\pi\rangle
+\sqrt{\frac{9}{16}}|\Delta,\eta\rangle
-\sqrt{\frac{1}{8}}|\Sigma^*,K\rangle
\nonumber \\
(+)~~&
|\mathbf{35};\frac{3}{2},1\rangle
&=
-\sqrt{\frac{1}{16}}|\Delta,\pi\rangle
+\sqrt{\frac{5}{16}}|\Delta,\eta\rangle
+\sqrt{\frac{5}{8}}|\Sigma^*,K\rangle
\end{eqnarray}
\begin{eqnarray}
(-)~~&
|\mathbf{27};2,0\rangle
&=
\phantom{+}\sqrt{\frac{3}{4}}|\Delta,{\bar K}\rangle
-\sqrt{\frac{1}{4}}|\Sigma^*,\pi\rangle
\nonumber \\
(+)~~&
|\mathbf{35};2,0\rangle
&=
\phantom{+}\sqrt{\frac{1}{4}}|\Delta,{\bar K}\rangle
+\sqrt{\frac{3}{4}}|\Sigma^*,\pi\rangle
\end{eqnarray}
\begin{eqnarray}
(+)~~&
|\mathbf{35};2,2\rangle
&=
\phantom{+}|\Delta,K\rangle
\end{eqnarray}
\begin{eqnarray}
(+)~~&
|\mathbf{35};\frac{5}{2},1\rangle
&=
\phantom{+}|\Delta,\pi\rangle
\end{eqnarray}

\end{widetext}

\end{document}